\documentclass[onecolumn,showpacs,showkeys,amsmath,amssymb,superscriptaddress,groupedaddress]{revtex4}
\usepackage{ifthen}
\newboolean{prd}
\setboolean{prd}{false}
\newboolean{arxiv}
\setboolean{arxiv}{true}
\newboolean{notprd}
\setboolean{notprd}{true}
\ifprd
\setboolean{notprd}{false}
\fi
\newboolean{notarxiv}
\setboolean{notarxiv}{true}
\ifarxiv
\setboolean{notarxiv}{false}
\fi
\usepackage{graphicx}
\usepackage[usenames]{xcolor}
\usepackage{subfigure}
\usepackage{amssymb}
\usepackage[latin1]{inputenc}
\usepackage{amsfonts}
\usepackage{setspace}
\usepackage{amsmath}
\usepackage{rotating}
\usepackage{ulem}
\usepackage{color}
\xdefinecolor{mylinkcolor}{rgb}{0,0,0.5}
\usepackage[
	bookmarksnumbered, bookmarksopen, bookmarksopenlevel=2,
	breaklinks=true,
	colorlinks=true, filecolor=mylinkcolor, citecolor=mylinkcolor,
	linkcolor=mylinkcolor, urlcolor=mylinkcolor, menucolor=mylinkcolor,
]{hyperref}

\usepackage{wasysym}
\newcommand{\be}{\begin{equation}}
\newcommand{\ee}{\end{equation}}
\newcommand{\bea}{\begin{eqnarray}}
\newcommand{\eea}{\end{eqnarray}}

\allowdisplaybreaks
\begin{document}

\hypersetup{
	pdftitle={},
	pdfauthor={}
}
%%%%%%%%%%%%%%%%%%%%%%%%%%%%%%%%%%%%%%%%%%%%%%%%%%%%%%%%%%%%%%%%%%%%%%%%%%%%%%%%%%%%%%%%%%%%%%%%%%%
%%%%%%%%%%%%%%%%%%%%%%%%%%%%%%%%%%%%%%%%%%%%%%%%%%%%%%%%%%%%%%%%%%%%%%%%%%%%%%%%%%%%%%%%%%%%%%%%%%%
%%%%%%%%%%%%%%%%%%%%%%%%%%%%%%%%%%%%%%%%%%%%%%%%%%%%%%%%%%%%%%%%%%%%%%%%%%%%%%%%%%%%%%%%%%%%%%%%%%%
\title{ Entropy of a box of gas in an external gravitational field --- revisited}
%\author{Sumanta Chakraborty}
%\email{sumanta@iucaa.in, sumantac.physics@gmail.com}
\author{Sourav Bhattacharya}
\email{sbhatta@iiitrpr.ac.in}
\affiliation{Department of Physics, Indian Institute of Technology Ropar, Rupnagar, Punjab 140 001, India}
\author{Sumanta Chakraborty}
\email{sumantac.physics@gmail.com}
\affiliation{Department of Theoretical Physics,
Indian Association for the Cultivation of Science, Kolkata 700 032, India}
\author{T.~Padmanabhan}
\email{paddy@iucaa.in}
\affiliation{IUCAA, Post Bag 4, Ganeshkhind, Pune University Campus, Pune 411 007, India}
\date{\today}
%%%%%%%%%%%%%%%%%%%%%%%%%%%%%%%%%%%%%%%%%%%%%%%%%%%%%%%%%%%%%%%%%%%%%%%%%%%%%%%%%%%%%%%%%%%%%%%%%%%
%%%%%%%%%%%%%%%%%%%%%%%%%%%%%%%%%%%%%%%%%%%%%%%%%%%%%%%%%%%%%%%%%%%%%%%%%%%%%%%%%%%%%%%%%%%%%%%%%%%
%%%%%%%%%%%%%%%%%%%%%%%%%%%%%%%%%%%%%%%%%%%%%%%%%%%%%%%%%%%%%%%%%%%%%%%%%%%%%%%%%%%%%%%%%%%%%%%%%%%
\begin{abstract}
\noindent
Earlier it was shown that the entropy of an ideal gas, contained in a box and moving in a gravitational field, develops an area dependence when it approaches the horizon of a static, spherically symmetric spacetime. Here we extend the above result in two directions; viz., to (a) the stationary axisymmteric spacetimes and (b) time dependent cosmological spacetimes evolving asymptotically to the de Sitter or the Schwarzschild de Sitter spacetimes. While our calculations are exact for the stationary axisymmetric spacetimes, for the cosmological case we present an analytical expression of the entropy when the spacetime is close to the de Sitter or the Schwarzschild de Sitter spacetime. Unlike the static spacetimes, there is no hypersurface orthogonal timelike Killing vector field in these cases. Nevertheless, the results hold and the entropy develops an area dependence in the appropriate limit.
\end{abstract}
%%%%%%%%%%%%%%%%%%%%%%%%%%%%%%%%%%%%%%%%%%%%%%%%%%%%%%%%%%%%%%%%%%%%%%%%%%%%%%%%%%%%%%%%%%%%%%%%%%%
%%%%%%%%%%%%%%%%%%%%%%%%%%%%%%%%%%%%%%%%%%%%%%%%%%%%%%%%%%%%%%%%%%%%%%%%%%%%%%%%%%%%%%%%%%%%%%%%%%%
%%%%%%%%%%%%%%%%%%%%%%%%%%%%%%%%%%%%%%%%%%%%%%%%%%%%%%%%%%%%%%%%%%%%%%%%%%%%%%%%%%%%%%%%%%%%%%%%%%%
\vskip .5cm
\noindent

\maketitle
%%%%%%%%%%%%%%%%%%%%%%%%%%%%%%%%%%%%%%%%%%%%%%%%%%%%%%%%%%%%%%%%%%%%%%%%%%%%%%%%%%%%%%%%%%%%%%%%%%%
%%%%%%%%%%%%%%%%%%%%%%%%%%%%%%%%%%%%%%%%%%%%%%%%%%%%%%%%%%%%%%%%%%%%%%%%%%%%%%%%%%%%%%%%%%%%%%%%%%%
%%%%%%%%%%%%%%%%%%%%%%%%%%%%%%%%%%%%%%%%%%%%%%%%%%%%%%%%%%%%%%%%%%%%%%%%%%%%%%%%%%%%%%%%%%%%%%%%%%%
\section{Introduction}\label{Intro}
\noindent
Killing horizons are null surfaces which act like one-way membranes, hiding information from observers who do not cross them. It is well known that such observers associate thermodynamic characteristics like entropy and temperature to these horizons (see e.g., Chapter~8 of~\cite{Padmanabhan}). The well known examples include the Schwarzschild and the Rindler spacetimes. In the Schwarzschild spacetime, an observer located at any constant radial coordinate $r$ such that $r>2M$, associates an entropy $S=A/4$ and a temperature $T=(1/8\pi M)$ with the black hole horizon located at $r=2M$~\cite{Bekenstein1,Bekenstein2,Bekenstein3,Hawking2}, where $A$ and $M$ are respectively the area of the event horizon and the black hole's mass. Similarly, an observer at constant spatial coordinate $x$ (with $x>0$) in a Rindler spacetime (with the metric $ds^2 = -g^2 x^2 dt^2 + d{\mathbf x}^2$) will associate a temperature $T=(g/2\pi)$ and entropy density $1/4$ with the Rindler horizon~\cite{Davies,Unruh:1976db}, located at 
$x=0$ (see also~\cite{Parker} for a review). On the other hand, the freely falling observers in either of these spacetimes will not associate any thermodynamic features with the horizon, because they have access to regions in both the interior and exterior of the event horizon (or, Rindler horizon) unlike the case of static observers (see~\cite{Padmanabhan1,Padmanabhan:2013nxa,Jacobson:2003wv} for details and also references therein) located outside the horizon. 

The fact that we may erect locally a flat coordinate chart around a given event of any non-singular spacetime, and that the flat chart could be mapped onto a Rindler frame, leads one to associate a local temperature and hence a local thermodynamic description of spacetimes~\cite{Padmanabhan1,Padmanabhan:2013nxa,Buchholz:2006iv}. This has paved the way towards an emergent perspective for gravity, since one can associate thermodynamic quantities locally at any spacetime event, pointing towards an underlying microscopic statistical description for gravity~\cite{Padmanabhan:2015zmr,Padmanabhan:2014jta, Padmanabhan:2010xe, Padmanabhan:2016eld}. This view has also received significant support from various other investigations, notably, one can introduce a thermodynamic perspective to the gravitational field equations near any null surface. In particular, various projections of the same can be written as the Navier-Stokes equation of fluid dynamics, as a thermodynamic identity as well as the heating and cooling of 
the spacetime \cite{Chakraborty:2015hna,Chakraborty:2015aja,Kolekar:2011gw,Padmanabhan:2010rp}. Remarkably enough, most of these thermodynamic properties carry over quite naturally (albeit non-trivially) to more general class of gravity theories \cite{Chakraborty:2014joa,Chakraborty:2014rga}. These results also have implications for classical gravity: for example,  until recently, it was not very clear how to formulate the variational principle for gravity when part of  the boundary is null~\cite{Padmanabhan:2014lwa,Parattu:2015gga,Parattu:2016trq,Chakraborty:2016yna}. A natural question that emerges in this context is the following: does the interpretation of $({\rm Area}/4)$ as the entropy still applies to an arbitrary null surface, which is not a Killing horizon? Surprisingly, it turns out that such existence of thermodynamic properties is not necessarily true only for Killing horizons --- in fact it was shown recently in~\cite{Chakraborty:2016dwb} using the Gaussian null coordinates that for a large 
class of generic null surfaces, which are {\it not} necessarily Killing horizons, the observers who do not cross them would still associate entropy densities $(1/4)$ with the relevant areas.

Given that the black holes have thermodynamic properties, it is natural to ask whether such notions could be associated with cosmological or Hubble horizons as well. Among them, perhaps the most interesting  one is that of the de Sitter space, which has a Killing horizon. The de Sitter horizon could also be associated with thermodynamic properties such as the entropy and temperature, qualitatively similar to that of the black hole \cite{Gibbons:1977mu,Traschen:1999zr,Choudhury:2004ph,Bhattacharya:2013tq,Bhattacharya:2015mja}. For cosmological spacetimes other than the de Sitter, the Hubble horizon, $r(t)=H^{-1}(t)$ (where $t$ is the comoving time, $r(t)$ is the proper radius and $H(t)$ is the Hubble rate) is \textit{not} a Killing horizon. For a spatially flat cosmology,  the Hubble horizon also coincides with the apparent horizon, where the expansion corresponding to one of the two principle null congruences vanishes while the other being positive. The notion of apparent horizon is dependent upon the 
spacetime foliation. Attempts to build thermodynamics by associating entropy and temperature with such horizons can be seen in, e.g. \cite{Cai:2005ra, Tian:2014sca, Saha:2012nz} and  in references therein. However, we note that unlike the Killing horizons, the Hubble or apparent horizon is {\it not} a null surface in general \cite{Faraoni:2015ula} as the normal to such surfaces, $\nabla_a (r(t) H(t))$, is not in general a null vector. 

In the usual statistical mechanics of normal matter, the entropy depends upon the volume of the system, instead of its area. Thus, the horizons are indeed some special objects -- qualitatively or quantitatively, as far as the thermodynamic properties are concerned. It was shown in~\cite{Padmanabhan:1998jp, Padmanabhan:1998vr} that in order for the entropy-area relation to hold, the density of states near the horizon must have an exponential `pile up' behaviour and any effective theory describing a quantum field near the horizon must be non-local, over the Planck length scale near the horizon. In~\cite{Kolekar:2010py}, the dynamics of a box containing an ideal gas moving in a static and spherically symmetric  black hole spacetime was considered and the area dependence of the entropy of the gas was demonstrated near the horizon by introducing the Planck length to be the cut off near the horizon. Such a cut off could be thought of as an ultraviolet cut off, making some near horizon divergent integrals finite. 
Similar result was obtained in a quasi-static  gravitational collapse scenario in~\cite{Kolekar:2011bb}, for both Einstein and Lanczos-Lovelock gravity. We further refer our reader to~\cite{Zerbini:1996zw, Frolov:1998vs, Solodukhin:2011gn} and references therein, for different perspective including entanglement of black hole entropy using matter fields as probes. 

As emphasized earlier, the entropy area relation in general relativity transcends the horizons which arise in static spacetimes and holds good in stationary spacetimes and  cosmological spacetimes. We already know from previous work that  the entropy for a \textit{normal} thermodynamic system also does scale as area for static spacetimes when it is close to the horizon. These two facts raise the following interesting question viz., whether the entropy of a normal thermodynamic system also scales as area if it is located near the stationary or cosmological horizon.  It is also of  interest to understand what happens to the entropy of a box of ideal gas when it is near the apparent horizon in a cosmological spacetime, which is \textit{not} a null surface. Given this motivational backdrop we will, in this work, extend the formalism developed in \cite{Kolekar:2010py} to spacetimes with a positive cosmological constant admitting cosmological or the Hubble horizon as well to rotating solutions in general 
relativity. First, we shall consider a box of ideal gas  moving in a general stationary axisymmetric spacetime endowed with a positive cosmological constant (e.g., the Kerr-Newman-de Sitter or the Plebanski-Demianski-de Sitter). We shall assume that the spacetime admits Killing horizons --- both cosmological and the black hole horizons. We shall demonstrate the area scaling of the gas's entropy for both the cases. Next we shall consider a box moving close to the Hubble horizon of a  cosmological spacetime with scale factor being appropriate for the $\Lambda$-cold dark matter ($\Lambda{\rm CDM}$ for short) or the $\Lambda$-radiation. We will show that, an area scaling of the entropy of the gas always exists and as the spacetime asymptotically evolves to the de Sitter, the entropy of the gas smoothly merges with the de Sitter Killing horizon behaviour. We also will present a generalization of this result with a central mass/black hole. 

The rest of the paper is organized as follows. In the next section we describe the necessary geometric framework and derive the entropy in stationary axisymmetric spacetimes. Sec.~3 describes the calculation of the entropy and the emergence of the area law for cosmological spacetimes. The conclusions are summarized  in Sec.~4. We shall assume mostly positive signature for the metric $(-,+,+,+)$ and would set $c=G=\kappa_{\rm B}=1$ throughout. We shall also assume spatially flat geometry ($k=0$), as the other plausible choice $k=+1$ seems to be relevant only for describing structures at scales much smaller than the Hubble horizon~\cite{Paddy}.
%%%%%%%%%%%%%%%%%%%%%%%%%%%%%%%%%%%%%%%%%%%%%%%%%%%%%%%%%%%%%%%%%%%%%%%%%%%%%%%%%%%%%%%%%%%%%%%%%%%
%%%%%%%%%%%%%%%%%%%%%%%%%%%%%%%%%%%%%%%%%%%%%%%%%%%%%%%%%%%%%%%%%%%%%%%%%%%%%%%%%%%%%%%%%%%%%%%%%%%
%%%%%%%%%%%%%%%%%%%%%%%%%%%%%%%%%%%%%%%%%%%%%%%%%%%%%%%%%%%%%%%%%%%%%%%%%%%%%%%%%%%%%%%%%%%%%%%%%%%
\section{Stationary axisymmetric spacetimes}
%%%%%%%%%%%%%%%%%%%%%%%%%%%%%%%%%%%%%%%%%%%%%%%%%%%%%%%%
%%%%%%%%%%%%%%%%%%%%%%%%%%%%%%%%%%%%%%%%%%%%%%%%%%%%%%%%
%%%%%%%%%%%%%%%%%%%%%%%%%%%%%%%%%%%%%%%%%%%%%%%%%%%%%%%%
\subsection{The metric and the coordinate systems}
\noindent
We consider a box of ideal gas moving in a stationary axisymmetric spacetime, endowed with a positive cosmological constant $\Lambda$. In order to model such a generic spacetime, let us first consider the example of the Kerr-Newman-de Sitter spacetime, representing a charged and rotating black hole sitting in the de Sitter universe (see e.g.~\cite{Akcay:2010vt} and also references therein) whose spacetime metric reads,
%%%%%%%%%%%%%%%%%%%%%%%%%%%%%%%%%%%%%%%%%%%%%%%%%%%%%%%%%%%%%%%%%%%%%%
\begin{eqnarray}
ds^2=-\frac{\Delta_r-a^2\sin^2\theta \Delta_{\theta}}{\rho^2}dt^2 -\frac{2a \sin^2 \theta }{\rho^2 \Xi} \left( (r^2+a^2 )\Delta_{\theta}-\Delta_r\right)dt d\phi \nonumber\\ + \frac{\sin^2 \theta }{\rho^2 \Xi^2} \left( (r^2 +a^2)^2 \Delta_{\theta}-\Delta_r a^2 \sin^2\theta\right)d\phi^2 + \frac{\rho^2}{\Delta_r}dr^2 + \frac{\rho^2}{\Delta_{\theta}}d\theta^2
\label{es1}
\end{eqnarray}
%%%%%%%%%%%%%%%%%%%%%%%%%%%%%%%%%%%%%%%%%%%%%%%%%%%%%%%%%%%%%%%%%%%%%%
where we have defined,
%%%%%%%%%%%%%%%%%%%%%%%%%%%%%%%%%%%%%%%%%%%%%%%%%%%%%%%%%%%%%%%%%%%%%%
\begin{eqnarray}
\Delta_r = (r^2 +a^2) (1-H_0^2 r^2 ) -2Mr  +Q^2, \quad \Delta_{\theta} =1 + H_0^2 a^2 \cos^2\theta,  \quad \Xi = 1+ H_0^2 a^2,  \quad \rho^2 =r^2 +a^2 \cos^2 \theta 
\label{es2}
\end{eqnarray}
%%%%%%%%%%%%%%%%%%%%%%%%%%%%%%%%%%%%%%%%%%%%%%%%%%%%%%%%%%%%%%%%%%%%%
with $H_0^2 =\Lambda/3$ and $M$, $Q$ and $a$ are respectively the parameters specifying mass, charge and angular momentum of the black hole. Setting $a=0$ recovers the static Reissner-N\"{o}rdstrom-de Sitter spacetime whereas setting further $M=0=Q$ recovers the de Sitter spacetime written in the static coordinate system. The cosmological and the black hole event horizons are respectively given by the largest (say, $r_C$) and the next to the largest (say, $r_H$) roots of $\Delta_r=0$.  

The Kerr-Newman-de Sitter spacetime is stationary and axisymmetric --- it is endowed with two Killing vector fields $\xi^a \equiv (\partial_t)^a$ and $\phi^a \equiv (\partial_{\phi})^a$, generating respectively, the stationarity and axisymmetry. These two vector fields, being coordinate fields, commute
%%%%%%%%%%%%%%%%%%%%%%%%%%%%%%%%%%%%%%%%%%%%%%%%%%%%%%%%%%%%%%%%%%%%%
\begin{eqnarray}
[\xi,~\phi]^a=\pounds_{\xi} \phi^a=0 
\label{es3}
\end{eqnarray}
%%%%%%%%%%%%%%%%%%%%%%%%%%%%%%%%%%%%%%%%%%%%%%%%%%%%%%%%%%%%%%%%%%%%%%
Since the 2-planes orthogonal to these two Killing vector fields are spanned by coordinate vector fields, $(\partial_r)^a$ and $(\partial_{\theta})^a$, they also commute with each other. This implies that any two vector fields spanning these 2-planes form a Lie algebra and hence the $r-\theta$ planes are integral sub-manifolds of the spacetime~\cite{Wald:1984rg}. In other words, when these planes are Lie-dragged along $\xi^a$ and/or $\phi^a$, they remain intact.

We note that, unlike in the case of the static spacetimes, the timelike Killing vector field is now not hypersurface orthogonal, as $\xi\cdot \phi = g_{t\phi} \neq 0$. The surface $\xi\cdot \xi = g_{tt}=0$ defines the ergosphere. Since both the black hole and the cosmological horizons correspond to $\Delta_r =0$, it is clear that unlike the static spacetimes, $\xi^a$ is spacelike on the horizons. This implies that unlike the $\Lambda \leq 0$ stationary axisymmetric spacetimes, we have two ergospheres, instead of one. From Eq.~(\ref{es2}) one can determine the locations of these ergospheres as --- (a) one surrounding the black hole and the other one appearing before the cosmological event horizon. These ergospheres intersect the respective horizons at the axial points, $\theta=0,~\pi$, where the effect of rotation vanishes. 

The angular velocities $\alpha= -(\xi\cdot \phi)/ (\phi \cdot \phi)=-g_{t\phi}/g_{\phi \phi}$ at the two horizons are respectively given by 
%%%%%%%%%%%%%%%%%%%%%%%%%%%%%%%%%%%%%%%%%%%%%%%%%%%%%%%%
\begin{eqnarray}
\alpha_H = -\frac{\xi\cdot \phi}{ \phi \cdot \phi}\Bigg\vert_{r=r_H}=\frac{a \Xi}{ r_H^2 +a^2}, \qquad  \alpha_C =    -\frac{\xi\cdot \phi}{ \phi \cdot \phi}\Bigg\vert_{r=r_C}= \frac{a \Xi}{ r_C^2 +a^2}.
\label{es4}
\end{eqnarray}
%%%%%%%%%%%%%%%%%%%%%%%%%%%%%%%%%%%%%%%%%%%%%%%%%%%%%%%%
The Killing vector fields that become null on the black hole and the cosmological horizons are respectively given by 
%%%%%%%%%%%%%%%%%%%%%%%%%%%%%%%%%%%%%%%%%%%%%%%%%%%%%%%%
\begin{eqnarray}
\chi_H^a = \xi^a +\alpha_H \phi^a, \qquad \chi_C^a =\xi^a + \alpha_C \phi^a,
\label{es5}
\end{eqnarray}
%%%%%%%%%%%%%%%%%%%%%%%%%%%%%%%%%%%%%%%%%%%%%%%%%%%%%%%
it is easy to see using the metric functions that $\chi_H \cdot \chi_H \,(r\to r_H)$ and $\chi_C \cdot \chi_C \,(r\to r_C)$ vanish as ${\cal O } (\Delta_r)$. This completes the basic geometric description of the Kerr-Newman-de Sitter spacetime, given by Eq.~(\ref{es1}).

Keeping this explicit example in mind, we shall now build a model for addressing general stationary axisymmetric spacetimes. Although we are primarily interested in $(3+1)$-dimensions, as we shall see at the end of this section that the results easily generalizes to arbitrary higher dimensions as well. More details of the following discussions can be seen in~\cite{Bhattacharya:2015mja} and references therein. We assume that the two Killing vector fields of the spacetime still commute (see for example Eq.~(\ref{es3})) and the 2-planes orthogonal to them are integral sub-manifolds. Note that this does not necessarily require that these vector fields are coordinate fields --- in fact it only means that we may locally define coordinates along those Killing vector fields.

We shall next construct a suitable $(3+1)$-foliation of the spacetime in between the two Killing horizons as follows: We define a vector field  $\chi^a= \xi^a -(\xi\cdot \phi)/(\phi\cdot \phi)\, \phi^a$ such that $\chi^a\phi_a$ identically vanishes everywhere. Also, $\chi\cdot \chi = \xi\cdot \xi - (\xi\cdot \phi)^2/(\phi\cdot \phi)=-{\tilde\beta}^2~({\rm say})$, so that $\chi^a$ is timelike whenever ${\tilde\beta}^2>0$. Since the term $(\xi\cdot \phi)^2/(\phi\cdot \phi)$ is positive definite, it is clear that $\chi^a$ is manifestly timelike outside the ergospheres. Its behaviour inside the ergosphere will be discussed in due course.

The fact that the 2-planes orthogonal to the Killing fields are integral sub-manifolds and hence any two vector fields tangent to them form a Lie algebra, implies following Frobenius-like conditions~\cite{Wald:1984rg},
%%%%%%%%%%%%%%%%%%%%%%%%%%%%%%%%%%%%%%%%%%%%%%%%%%%%%%%%%%%%%
\begin{eqnarray}
\xi_{[a} \phi_b \nabla_c \phi_{d]}=0= \phi_{[a}\xi_b\nabla_c\xi_{d]}
\label{es6}
\end{eqnarray}
%%%%%%%%%%%%%%%%%%%%%%%%%%%%%%%%%%%%%%%%%%%%%%%%%%%%%%%%%%%%
which can be rewritten in terms of our vector field $\chi^a$,
%%%%%%%%%%%%%%%%%%%%%%%%%%%%%%%%%%%%%%%%%%%%%%%%%%%%%%%%%%%%
\begin{eqnarray}
\chi_{[a} \phi_b \nabla_c \phi_{d]}=0= \phi_{[a}\chi_b\nabla_c\chi_{d]}
\label{es7}
\end{eqnarray}
%%%%%%%%%%%%%%%%%%%%%%%%%%%%%%%%%%%%%%%%%%%%%%%%%%%%%%%%%%%%
Denoting the function $-(\xi\cdot \phi)/(\phi\cdot \phi) $ by $\alpha (x)$, we have
%%%%%%%%%%%%%%%%%%%%%%%%%%%%%%%%%%%%%%%%%%%%%%%%%%%%%%%%%%%%
\begin{eqnarray}
\nabla_{a}\chi_{b}+\nabla_b\chi_a = \phi_a\nabla_b \alpha +\phi_b \nabla_a\alpha
\label{es8}
\end{eqnarray}
%%%%%%%%%%%%%%%%%%%%%%%%%%%%%%%%%%%%%%%%%%%%%%%%%%%%%%%%%%%
We also get, from the commutativity of the two Killing vector fields,
%%%%%%%%%%%%%%%%%%%%%%%%%%%%%%%%%%%%%%%%%%%%%%%%%%%%%%%%%%%
\begin{eqnarray}
\pounds_{\phi}\chi^a =0 \qquad {\rm and }~~~~~ \pounds_{\chi}\alpha =0=\pounds_{\phi}\alpha
\label{es9}
\end{eqnarray}
%%%%%%%%%%%%%%%%%%%%%%%%%%%%%%%%%%%%%%%%%%%%%%%%%%%%%%%%%%%%%
Setting $\alpha=0$ one immediately recovers the static spacetime. Using Eqs.~(\ref{es8}) and (\ref{es9}), we can derive from the second of Eqs.~(\ref{es7}),
%%%%%%%%%%%%%%%%%%%%%%%%%%%%%%%%%%%%%%%%%%%%%%%%%%%%%%%%
\begin{eqnarray}
\nabla_{[a}\chi_{b]}= 2{\tilde\beta}^{-2}\left(\chi_b \nabla_a{\tilde\beta}^2 -\chi_a\nabla_b{\tilde \beta}^2 \right)
\label{es10}
\end{eqnarray}
%%%%%%%%%%%%%%%%%%%%%%%%%%%%%%%%%%%%%%%%%%%%%%%%%%%%%%%%%
which implies that $\chi_a$ satisfies the Frobenius condition of hypersurface orthogonality, i.e., $\chi_{[a}\nabla_b\chi_{c]}=0$. In other words, the vector field $\chi^a$ is orthogonal to the family of spacelike hypersurfaces spanned by $\phi^a$ and the aforementioned 2-sub-manifolds. We also note that the price we have paid in making this foliation is that, $\chi^a$ is not a Killing vector field now (see Eq.~(\ref{es8})) unlike the static spacetimes. The metric now reads
%%%%%%%%%%%%%%%%%%%%%%%%%%%%%%%%%%%%%%%%%%%%%%%%%%%%%%%%%
\begin{eqnarray}
g_{ab}=- {\tilde\beta}^{-2}\chi_a\chi_b +h_{ab}
\label{es11'}
\end{eqnarray}
%%%%%%%%%%%%%%%%%%%%%%%%%%%%%%%%%%%%%%%%%%%%%%%%%%%%%%%%%%
Note that since the foliation field $\chi ^{a}$ can be thought of as orthogonal to $t=\textrm{constant}$ hypersurfaces,  $h_{ab}$ is the induced 3-metric over the spatial hypersurfaces orthogonal to $\chi^a$, spanned, e.g. in the Boyer-Lindquist coordinates, by $(r,\theta,\phi)$.

As we have seen earlier, $\chi^a$ is manifestly timelike outside the ergospheres. Since within the ergospheres, $\xi^a$ is spacelike ($\xi\cdot \xi > 0$), we might expect that $\chi_a\chi^a=-{\tilde\beta}^2$ would be vanishing at some point. Eq.~(\ref{es10}) shows that {\it on such a surface} using the torsion free condition, we must have $\chi_{[b}\nabla_{a]}{\tilde\beta}^2 = {\tilde\beta}^2\partial_{[a}\chi_{b]}=0$. In other words, the normal $\nabla_a{\tilde\beta}^2$ to any ${\tilde\beta}^2=0$ surface becomes parallel to the foliation field, $\chi_a$,
%%%%%%%%%%%%%%%%%%%%%%%%%%%%%%%%%%%%%%%%%%%%%%%%%%%%%%%%%
\begin{eqnarray}
\nabla_a{\tilde\beta}^2= -2\kappa(x) \chi_a,
\label{es11}
\end{eqnarray}
%%%%%%%%%%%%%%%%%%%%%%%%%%%%%%%%%%%%%%%%%%%%%%%%%%%%%%%%%%
where $\kappa(x)$ is a function, assumed to be smooth. It turns out from the above equation that $\pounds_{\chi} \kappa (x) =0$. Then, if $s$ is a parameter along $\chi^a$  (i.e., $\chi^a\nabla_a s := 1$), it is easy to see that the vector field $k^a := e^{-\kappa (x) s} \chi^a$ is a null geodesic. By considering the Raychaudhuri equation for the null congruence $\{k^a\}$, one readily arrives at (see~\cite{Bhattacharya:2015mja} and references therein), 
%%%%%%%%%%%%%%%%%%%%%%%%%%%%%%%%%%%%%%%%%%%%%%%%%%%%%%%%%%%
\begin{eqnarray}
T_{ab}k^ak^b + \frac{f^2 e^{-2\kappa s}}{2} (D_a \alpha)(D^a \alpha)=0,
\label{es12}
\end{eqnarray}
%%%%%%%%%%%%%%%%%%%%%%%%%%%%%%%%%%%%%%%%%%%%%%%%%%%%%%%%%%%%%%
where we have denoted $\phi_a\phi^a =f^2$ and $D_a$ is the derivative operator tangent to the ${\tilde\beta}^2=0$ hypersurface. Since $\pounds_{\chi}\alpha=0$ always, the inner product $(D_a\alpha)(D^a\alpha)$ is positive definite. It is reasonable to assume that the matter energy-momentum tensor satisfies the weak and null energy condition, implying $T_{ab}k^ak^b\geq 0$. Then the left hand side of Eq.~(\ref{es12}) consists of positive definite quantities and the vanishing of their sum implies that each of them are vanishing. This means that the function $\alpha (x)$ is a constant over any ${\tilde\beta}^2=0$ hypersurface. In other words, the hypersurface orthogonal  vector field $\chi^a=\xi^a+\alpha(x)\phi^a$ becomes Killing whenever it is null and that null hypersurface is a Killing horizon. This result is analogous to the explicit result of the Kerr-Newman-de Sitter spacetime discussed above. However, we note that the above general result has been proven without assuming {\it any} particular functional 
form of the norm of the various basis vector fields, nor assuming any particular matter field other than imposing the generic energy conditions. In other words, we have found a well behaved timelike vector field in a general stationary axisymmetric spacetime, foliating the region between the two Killing horizons and becoming smoothly the Killing fields on the horizons. This will be useful for our future purpose.    

It then follows that the function $\kappa$, known as the surface gravity, is a constant on the Killing horizons~\cite{Wald:1984rg}. We shall assume that $\kappa \neq 0$. We have explicitly specified two basis vectors so far -- $\chi^a$ and $\phi^a$. For our purpose, we shall now specify another one, which would in particular be useful while dealing with the near horizons' geometries. Let us consider the vector field $R_a=\frac{1}{\kappa}\nabla_a{\tilde\beta}$. Using the commutativity of the two Killing vector fields, it is easy to see that $\chi^a\nabla_a{\tilde\beta}=0=\phi^a\nabla_a{\tilde\beta}$. By Eq.~(\ref{es11}) we have 
%%%%%%%%%%%%%%%%%%%%%%%%%%%%%%%%%%%%%%%%%%%%%%%%%%%%%%%%%%%%%%%%%%%
\begin{eqnarray}
\kappa^2 = \lim_{{\tilde\beta}^2\to 0} \frac{(\nabla_a {\tilde\beta}^2)(\nabla^a{\tilde\beta}^2)}{4{\tilde\beta}^2}\equiv \lim_{{\tilde\beta}^2\to 0}(\nabla_a{\tilde\beta})(\nabla^a {\tilde\beta})
\label{es13}
\end{eqnarray}
%%%%%%%%%%%%%%%%%%%%%%%%%%%%%%%%%%%%%%%%%%%%%%%%%%%%%%%%%%%%%%%%%
Thus, if $R$ is parameter along $R^a$ (such that $R^a\nabla_a R :=1$ and if $\lambda(x)$ is any function, $R^a\nabla_a \lambda=\frac{d\lambda}{dR}$), then
%%%%%%%%%%%%%%%%%%%%%%%%%%%%%%%%%%%%%%%%%%%%%%%%%%%%%%%%%%%%%%%
\begin{eqnarray}
R^aR_a= \frac{1}{\kappa}R^a\nabla_a {\tilde\beta} =\frac{1}{\kappa}\frac{d {\tilde\beta}}{dR}=1
\label{es14}
\end{eqnarray}
%%%%%%%%%%%%%%%%%%%%%%%%%%%%%%%%%%%%%%%%%%%%%%%%%%%%%%%%%%%%%
which means, since $\kappa$ is a constant,
%%%%%%%%%%%%%%%%%%%%%%%%%%%%%%%%%%%%%%%%%%%%%%%%%%%%%%%%%%%%%
\begin{eqnarray}
R=\frac{{\tilde\beta}}{\kappa}
\label{es15}
\end{eqnarray}
%%%%%%%%%%%%%%%%%%%%%%%%%%%%%%%%%%%%%%%%%%%%%%%%%%%%%%%%%%%%%%%%
On the other hand, if $\mu^a$ is the fourth basis tangent to the horizon, we must have $\mu^a R_a \sim {\cal O}({\tilde\beta})  $ infinitesimally close to the Killing horizon, as ${\tilde\beta} = 0$ on the horizon. Thus, with our specified basis, we have ${\tilde\beta}^2=\kappa^2 R^2$ and the metric in Eq.~(\ref{es11'}) near any of the Killing horizons could be written as
%%%%%%%%%%%%%%%%%%%%%%%%%%%%%%%%%%%%%%%%%%%%%%%%%%%%%%%%%%%%%%%%%
\begin{eqnarray}
g_{ab}= -(\kappa R)^{-2}\chi_a\chi_b +R_a R_b+ f^{-2}\phi_a\phi_b + \mu^{-2}\mu_a\mu_b
\label{es16}
\end{eqnarray}
%%%%%%%%%%%%%%%%%%%%%%%%%%%%%%%%%%%%%%%%%%%%%%%%%%%%%%%%%%%%%%%%%%
where $\mu^2 =\mu_a\mu^a$. The $\chi-R$ part of the metric could now be identified with the Rindler space. Note that the square root of the norm ${\tilde\beta}$ has been used here as a coordinate along $R^a$, to express the notion of `off the horizon' direction. This would be  useful in defining the ultraviolet cut off while probing the near horizon geometry of a stationary axisymmetric spacetime.

Even though we have emphasized on $\Lambda>0$, the above analysis holds for anti-de Sitter or asymptotically flat spacetimes, for which the cosmological event horizon would be absent. 
We once again emphasize here that we have arrived at Eq.~(\ref{es16}) using purely generic arguments and assumptions, incorporating the effect of any matter field satisfying the weak/null energy conditions. 

As an explicit example, let us go back to the Kerr-Newman-de Sitter spacetime~(\ref{es1}), for which we have for the Killing horizons,
%%%%%%%%%%%%%%%%%%%%%%%%%%%%%%%%%%%%%%%%%%%%%%%%%%%%%%%%%%%%%%%%%
\begin{eqnarray}
-{\tilde\beta}^2 &=& g_{tt}-\frac{ (g_{t\phi})^2 }{g_{\phi\phi}}= -\frac{\Delta_r \rho^2}{(r^2+a^2)^2} +{\cal O}(\Delta_r^2)
\nonumber
\\
\kappa_{H,C}&=& \frac{\Delta'_r}{2(r^2+a^2)}\Bigg\vert_{r_H,r_C}=\frac{r_{H,C}(1-H_0^2r_{H,C}^2)-H_0^2 r_{H,C} (r_{H,C}^2+a^2)-M}{r^2_{H,C}+a^2}\nonumber\\
\label{es17}
\end{eqnarray}
%%%%%%%%%%%%%%%%%%%%%%%%%%%%%%%%%%%%%%%%%%%%%%%%%%%%%%%%%%%%%%%%%%
where the subscript $(H,C)$ denotes respectively, the black hole and the cosmological event horizon and the `prime' denotes differentiation with respect to the radial coordinate. The aforementioned basis vector $\mu^a$ equals $(\partial_{\theta})^a$ here and $(\partial_{\theta})^aR_a ={\cal O}(\sqrt{\Delta_r})$. In other words, near the horizons, the four vector fields $\chi^a, ~R^a,~\phi^a,~(\partial_{\theta})^a $ furnish a well behaved basis for the near horizon geometry.

Although we built our general formalism starting with the Kerr-Newman-de Sitter, it applies well to more general de Sitter black hole spacetimes as well. Such generalizations becomes necessary, when we recall the potential  non-uniqueness properties of de Sitter black holes~\cite{Boucher:1983cv}. For example, the general family of the Plebanski-Demianski-de Sitter class spacetimes has the metric~\cite{Griffiths:2005qp},
%%%%%%%%%%%%%%%%%%%%%%%%%%%%%%%%%%%%%%%%%%%%%%%%%%%%%%%%%%%%%%%%%
\begin{eqnarray}
ds^2=\frac{1}{\Omega^2}\left[-\frac{\Delta_r}{\rho^2}\left(dt-\left(a\sin^2\theta+4l\sin^2\frac{\theta}{2}\right)d\phi \right)^2+\frac{\rho^2}{\Delta_r}dr^2  + \frac{P}{\rho^2} \left(adt-\left(r^2 +(a+l)^2 \right) d\phi \right)^2      +\frac{\rho^2}{P}\sin^2\theta d\theta^2 \right]
\label{es18}
\end{eqnarray} 
%%%%%%%%%%%%%%%%%%%%%%%%%%%%%%%%%%%%%%%%%%%%%%%%%%%%%%%%%%%%%%%%%%%%%%
where
%%%%%%%%%%%%%%%%%%%%%%%%%%%%%%%%%%%%%%%%%%%%%%%%%%%%%%%%%%%%%%%%%%%%%%%%
\begin{eqnarray}
\Omega&=&1-\frac{\tilde{\alpha}}{\omega}\left( l+a\cos\theta\right)r, \quad \rho^2=r^2+\left( l+a\cos\theta\right)^2, \quad%\nonumber\\
P=\sin^2\theta \left(1-a_3\cos\theta-a_4\cos^2\theta\right)\nonumber\\
\Delta_r&=&\left(\omega^2 k+q^2+q_m^2\right)-2Mr+\epsilon r^2 -\frac{2\tilde{\alpha} n}{\omega}r^3-\left(\tilde{\alpha}^2 k+ H_0^2\right)r^4,
\label{es19}
\end{eqnarray} 
%%%%%%%%%%%%%%%%%%%%%%%%%%%%%%%%%%%%%%%%%%%%%%%%%%%%%%%%%%%%%%%%%%%%
The parameters $\tilde{\alpha}$, $\omega$, $q$, $q_m$, $\epsilon$ and 
$k$ are independent, and $a_3$ and $a_4$ are determined from them via some constraints. Physical meaning to these parameters could be asserted to only certain special subclasses of Eq.~(\ref{es18}). 
In particular, for $\tilde{\alpha}=0$, the above metric reduces to the Kerr-Newman-NUT-de Sitter solution~\cite{Griffiths:2005qp},
%%%%%%%%%%%%%%%%%%%%%%%%%%%%%%%%%%%%%%%%%%%%%%%%%%%%%%%%%%%%%%%%
\begin{eqnarray}
ds^2=-\frac{\Delta_r}{\rho^2}\left[dt-(a\sin^2\theta+4l \sin^2\theta/2)d\phi\right]^2+\frac{\rho^2}{\Delta_r}dr^2
+\frac{P}{\rho^2}\left[adt-(r^2+(a+l)^2)d\phi\right]^2+\frac{\rho^2}{P}\sin^2\theta d\theta^2,
\label{es20}
\end{eqnarray} 
%%%%%%%%%%%%%%%%%%%%%%%%%%%%%%%%%%%%%%%%%%%%%%%%%%%%%%%%%%%%%%%%%%%%
where
%%%%%%%%%%%%%%%%%%%%%%%%%%%%%%%%%%%%%%%%%%%%%%%%%%%%%%%%%%%%%%%%%%%%%
\begin{eqnarray}
\rho^2&=&r^2+\left( l+a\cos\theta\right)^2, \quad%\nonumber\\
P=\sin^2\theta \left(1+ {4 a l H_0^2 \cos\theta}+ {H_0^2 a^2 \cos^2\theta}\right)\nonumber\\
\Delta_r&=&\left(a^2-l^2+q^2+q_m^2\right)-2Mr+r^2 -3H_0^2\left((a^2-l^2)l^2+(a^2/3+2l^2)r^2+ r^4/3\right),
\label{es21}
\end{eqnarray} 
%%%%%%%%%%%%%%%%%%%%%%%%%%%%%%%%%%%%%%%%%%%%%%%%%%%%%%%%%%%%%%%%%%%
where $q$ and $q_m$ are respectively electric and magnetic charges and $l$ is the Newman-Unti-Tamburino (NUT) parameter.
Note that, in order to have well-behaved black hole solution with $H_0=0$,  one must have both the acceleration parameter and the NUT charge to be vanishing. Nevertheless, in what follows we shall formally consider  the most general Plebanski-Demianski-de Sitter class given by Eq.~(\ref{es18}), assuming {\it implicitly} it indeed represents de Sitter black holes subject to suitable parameter values which, for our current purpose is not an explicit concern.

The black hole and the cosmological horizons are, as earlier given by the two largest roots of $\Delta_r=0$. The integrability of the 2-sub-manifolds orthogonal to the two commuting Killing vector fields follows trivially. The hypersurface orthogonal vector field $\chi^a$ behaves near the horizons as,
%%%%%%%%%%%%%%%%%%%%%%%%%%%%%%%%%%%%%%%%%%%%%%%%%%%%%%%%%%%%%%%%%%%%
\begin{eqnarray}
\chi_{H,C}&=& \xi^a +\frac{a}{r^2_{H,C}+(a+l)^2}\phi^a\nonumber\\
\chi^a\chi_a&=&-{\tilde\beta}^2=-\frac{\Delta_r\rho^2}{\Omega^2 [r^2+(a+l)^2]^2 } +{\cal O}(\Delta_r^2).
\label{es22}
\end{eqnarray} 
%%%%%%%%%%%%%%%%%%%%%%%%%%%%%%%%%%%%%%%%%%%%%%%%%%%%%%%%%%%%%%%%%%%%%
The above geometrical set up can also be generalized to higher dimensional spacetimes as well, permitting larger number of commuting Killing vector fields whose orthogonal space is spanned by integral sub-manifolds of appropriate dimensions.  In that case the hypersurface orthogonal timelike vector field $\chi^a$ is given by $\chi^a=\xi^a + \alpha_{(i)}\phi^a_{(i)}$ where $i$ denotes the number of axisymmetric Killing vector fields and the functions $\alpha_{(i)}$'s are determined by solving the algebraic equations, $\chi_a\phi^a_{(i)}=0$ for each $i$. In fact all existing stationary and axisymmetric solutions fall within the scope of this geometric set up. We shall not go into further details  here, and refer an interested reader to~\cite{Bhattacharya:2015mja} and references therein.  

Previous work shows that~\cite{Kolekar:2010py} it is necessary to introduce a cut-off near the Killing horizons to regularize some divergent  integrals in calculating the entropy. Such a cut off could also be physically justified from the fact that for any finite Killing-coordinate time interval, no particle can reach the horizon. The natural cut off in probing the near horizon geometry is the Planck length. Specifically, for a static and spherically symmetric non-extremal black hole spacetime,
$$ds^2=-f(r)dt^2+f^{-1}(r)dr^2+r^2d\Omega^2$$
we introduce the Planck length cut-off as  the minimum proper radial distance of the approaching side of the box 
%%%%%%%%%%%%%%%%%%%%%%%%%%%%%%%%%%%%%%%%%%%%%%%%%%%%%%%%%%%%%%%
\begin{eqnarray}
L_P:= \int_{r=r_H}^{r=L_a+r_H} \frac{dr}{\sqrt{f(r)}} \approx \sqrt{\frac {2L_a}{\kappa_H} }.
\label{es23}
\end{eqnarray} 
%%%%%%%%%%%%%%%%%%%%%%%%%%%%%%%%%%%%%%%%%%%%%%%%%%%%%%%%%%%%%%%%%%
where it is assumed that $L_a\ll r_H$, so that we could expand in the integrand, $f(r\to r_H)\approx 2\kappa_H(r-r_H)$. If we try to define a similar cut off for the stationary axisymmetric spacetimes, we face an immediate difficulty because the pure radial path near the horizon of a stationary axisymmetric spacetime makes no sense. In particular, within the ergosphere, all particle will rotate due to the frame dragging effect~\cite{Wald:1984rg}. Also, since the metric functions now depend upon both radial and the polar coordinate, defining proper length as above seems to be inconsistent.  For example, in the case of Eq.~(\ref{es1}), the proper radial distance should be $\int dr \sqrt{\frac{r^2+a^2\cos^2\theta}{\Delta_r}}$, which is $\theta$ dependent. 

To handle  this difficulty, we would instead use the coordinate $R$ defined in Eq.~(\ref{es15})
for our purpose. (Note that since ${\tilde\beta}$ is dimensionless and the surface gravity $\kappa$ has inverse length dimension; so $R$ has the dimension of length.) Since $\tilde{\beta}=0$ on the horizons, $R=0$ there, too and we simply have, in this new coordinate, 
%%%%%%%%%%%%%%%%%%%%%%%%%%%%%%%%%%%%%%%%%%%%%%%%%%%%%%%%%%%%%%%%%
\begin{eqnarray}
L_P:= \int_{0}^{L_a} dR.
\label{es24}
\end{eqnarray} 
%%%%%%%%%%%%%%%%%%%%%%%%%%%%%%%%%%%%%%%%%%%%%%%%%%%%%%%%%%%%%%%%%%
In other words, since $R\sim \tilde{\beta}$ and sufficiently close to the horizon $\tilde{\beta}$ must be monotonic, we are using the value of $\tilde{\beta}$ to indicate a measure of how far off we are from  the horizon. This does not depend upon any specific path. Thus the above generalization provides a natural way to impose a cut off in probing the near horizon geometry, for stationary axisymmetric spacetimes.
  
Before we end this discussion, let us illustrate how this works out explicitly for the Kerr-Newman-de Sitter spacetime, presented in  Eq.~(\ref{es1}). We consider the volume integral over the $(r,~\theta,~\phi)$-hypersurface,
%%%%%%%%%%%%%%%%%%%%%%%%%%%%%%%%%%%%%%%%%%%%%%%%%%%%%%%%%%%%%%%%%
\begin{eqnarray}
\int \sqrt{g_{\theta\theta}g_{\phi\phi}}d\theta d\phi \frac{\rho}{\sqrt{\Delta_r}}dr,
\label{es25}
\end{eqnarray} 
%%%%%%%%%%%%%%%%%%%%%%%%%%%%%%%%%%%%%%%%%%%%%%%%%%%%%%%%%%%%%%%%%
and take the near horizon limit, $\Delta_r \to 0 $. Taking the differential of Eq.~(\ref{es15}), we get
%%%%%%%%%%%%%%%%%%%%%%%%%%%%%%%%%%%%%%%%%%%%%%%%%%%%%%%%%%%%%%%%%
\begin{eqnarray}
dR =\frac{\rho \Delta_r' dr}{2(r^2+a^2) \sqrt{\Delta_r}\kappa }-\frac{\sqrt{\Delta_r}}{\kappa (r^2+a^2)}\left(\frac{2\rho r }{r^2+a^2}dr + \frac{a^2 \sin2\theta}{2\rho}d\theta\right) \approx \frac{\rho \Delta_r' dr}{2(r^2+a^2) \sqrt{\Delta_r}\kappa } = \frac{\rho dr }{\sqrt{\Delta_r}}
\label{es26}
\end{eqnarray} 
%%%%%%%%%%%%%%%%%%%%%%%%%%%%%%%%%%%%%%%%%%%%%%%%%%%%%%%%%%%%%%%%%%%
where in the last equality we have used the second of Eqs.~(\ref{es17}) and also have used the fact that since $\theta$ is tangent to the horizon, $d\theta$ really remains infinitesimal there. In other words, in the near
horizon limit we have 
%%%%%%%%%%%%%%%%%%%%%%%%%%%%%%%%%%%%%%%%%%%%%%%%%%%%%%%%%%%%%%%%%%%%%
\begin{eqnarray}
\int \sqrt{g_{\theta\theta}g_{\phi\phi}}d\theta d\phi \frac{\rho}{\sqrt{\Delta_r}}dr\, \to \int \sqrt{g_{\theta\theta}g_{\phi\phi}}d\theta d\phi dR,
\label{es27}
\end{eqnarray} 
%%%%%%%%%%%%%%%%%%%%%%%%%%%%%%%%%%%%%%%%%%%%%%%%%%%%%%%%%%%%%%%%%%%%%
which is exactly analogous to the static horizons written in Rindler coordinates, thereby establishing the universality of the horizon properties of these two kind of spacetimes. With these necessary ingredients, we are now ready to calculate the entropy of a box of gas.
%%%%%%%%%%%%%%%%%%%%%%%%%%%%%%%%%%%%%%%%%%%%%%%%%%%%%%%%
%%%%%%%%%%%%%%%%%%%%%%%%%%%%%%%%%%%%%%%%%%%%%%%%%%%%%%%%
%%%%%%%%%%%%%%%%%%%%%%%%%%%%%%%%%%%%%%%%%%%%%%%%%%%%%%%%
\subsection{Entropy calculation}
\noindent
Let us consider a box of ideal gas containing $N$ particles in thermal equilibrium with its surroundings at temperature $\beta^{-1}$. For a static spacetime, one can use  the canonical ensemble~\cite{Kolekar:2010py}. Let us denote the density of states and the phase space volume (respectively) by  $g(E)$ and $P(E)$. Then we have
\begin{eqnarray}
g(E)=\frac{dP(E)}{dE} \qquad Q(\beta) = \int e^{-\beta E} g(E) dE=   \int e^{-\beta E} dP(E)
\label{e15}
\end{eqnarray}
where $Q(\beta)$ is the canonical one-particle partition function, related to the density of states via a Laplace transform. For general static spacetimes, the phase space volume can be written as~\cite{Kolekar:2010py, Paddy2}  
%%%%%%%%%%%%%%%%%%%%%%%%%%%%%%%%%%
\begin{eqnarray}
P(E)=\int d^3x d^3p \,\Theta (E-|\xi^ap_a|)
\label{e16}
\end{eqnarray}
%%%%%%%%%%%%%%%%%%%%%%%%%%%%%%%%%%%%%%
where the integration is over any spacelike hypersurface orthogonal to the timelike Killing vector field, $\xi^a$. The $4$-momentum of a timelike or null geodesic is denoted by $p^{a}$ and $E=-\xi^ap_a$ is the conserved energy along the geodesic. By the definition of step function it is clear that the above integral essentially computes the total phase space volume \textit{within} the surface $E=|\xi ^{a}p_{a}|$. Further, the integration measure can be seen to be invariant under local Lorentz transformations -- the spatial volume gets contracted whereas the spatial momentum gets dilated -- effectively leaving the phase space measure invariant (a proof of the same can be found in standard textbooks like e.g., \cite{MTW,DeWitt,LL2}). A quick proof can be given using the invariance of $d^3p/E_p$ and $d^4x$ and using the fact the $dt/ds\sim E$ where the symbol $\sim$ stands for ``transforms as''. Then $(d^3p/E_p)(d^4x)(ds)\sim(d^3p/E_p)(d^3x)(dt/ds)\sim d^3pd^3x$. Since $(d^3p/E_p)(d^4x)(ds)$ is manifestly 
invariant, so is $d^3pd^3x$.

In a non-static spacetime, the definition of energy as a conserved quantity, with respect to a hypersurface orthogonal timelike Killing vector field does not exist. Nevertheless the quantity $\Theta (E+\xi^ap_a)$ could still be used anyway as follows. Let us replace $E$ with some scalar function $E(x)$ ($x$ can stand for all spatial and the temporal coordinates) and we take $E(x)=-\chi^ap_a$, where $\chi^a$ is some timelike vector field orthogonal to a family of 
spatial hypersurfaces used in the integration measure. If we take both $\chi^a$ and $p^a$ to be future directed (i.e., both $\chi^0,~p^0>0$~\cite{Wald:1984rg}), it is obvious that $E(x)$ would be positive. This is a formal argument in favour of using Eq.~(\ref{e16}) in general non-static spacetimes. Thus the above definition of the probability distribution for general spacetimes seems to be dependent on the foliation. Then it is natural to ask, what is the appropriate foliation to work with? The answer to this question lies in the notion of local temperature. Given a spacetime, only a particular foliation of it would lead to a correct value for the surface gravity in the Rindler limit and one must use this foliation for getting meaningful results, pertaining spacetime thermodynamics. While for static spacetimes with Killing horizons there is no such ambiguity -- one must always use the Killing time to define the foliation --- for non-static spacetimes, we shall use only those which ensure the existence of 
the proper Rindler limit of the spacetime.

Let us evaluate $P(E)$ for the stationary axisymmetric spacetimes, taking $\chi^a$ to be the foliation vector field, as discussed in the previous section. Since the gas molecules individually moves along geodesics,  we note that the quantities $\epsilon=-\xi_ap^a$ and $\lambda=\phi_ap^a$ are conserved, respectively being the energy and the orbital angular momentum. This means that $E(x)=-p_a\chi^a=(\epsilon-\lambda \alpha(x))$. Since $\alpha(x)$ is constant only on  horizons,  $E$ is not  conserved everywhere. Setting $\alpha=0$ recovers the static spacetime. We have from Eq.~(\ref{es11'}), for a timelike geodesic,  
\begin{eqnarray}
-\frac{(\chi\cdot p)^2}{\tilde{\beta}^2} + h_{ab}p^ap^b=-m^2
\label{es28}
\end{eqnarray}
But for the spatial momentum, we have clearly, $p^2\equiv h_{ab}p^ap^b$, where $p$ denotes the magnitude square of the spatial momentum, giving 
\begin{eqnarray}
p^2=\frac{(\epsilon-\lambda \alpha (x))^2}{\tilde{\beta^2}}-m^2
\label{es29}
\end{eqnarray}
Thus we have 
\begin{eqnarray}
P(\epsilon;\lambda)=\frac{4\pi}{3}\int \sqrt{h} d^3x \left[  \frac{(\epsilon-\lambda \alpha (x))^2}{\tilde{\beta^2}}-m^2 \right]^{\frac32} 
\label{es30}
\end{eqnarray}
To arrive at the above integral we have used the fact that the phase space volume includes a step function and hence the only contribution to the phase space volume will come from the interior of the surface $E=|\xi ^{a}p_{a}|$, which corresponds to a sphere in the momentum space with radius given by Eq.~(\ref{es29}). Thus in addition to a $\sqrt{h}$ factor, the momentum integral over the phase space simply yields $(4\pi/3)p^{3}$, where $|p|=\sqrt{h_{ab}p^{a}p^{b}}$. The above integral cannot be evaluated in general, even in the Schwarzschild spacetime. Nevertheless, apparently there is no reason for the above integral to lead to any area scaling of entropy, in general.

Let us evaluate the above integral, infinitesimally close to any of the horizons of a stationary axisymmetric spacetime. We use now Eq.~(\ref{es16}), $\alpha={\rm const.}$; the mass term goes away giving 
\begin{eqnarray}
P(\epsilon;\lambda)=\frac{4\pi (\epsilon - \lambda \alpha)^3}{3}\int [d^2X] \frac{dR}{\kappa^3 R^3}  
\label{esn30}
\end{eqnarray}
where $[d^2X]$ is the invariant `volume' measure of the spatial 2-surface on the horizon (spanned by $\mu^a$ and $\phi^a$ in Eq.~(\ref{es16}); for general compact horizons such as that of the Kerr-Newman-de Sitter, $\mu^a=(\partial_{\theta})^a$). Using the Planck length to be the ultraviolet cut off, we can perform the above integration. The side of the box away from the horizon gives very little contribution to the integral (this is in accordance with the exponential `pile up' of the accessible microstates near a Killing horizon~\cite{Padmanabhan:1998jp, Padmanabhan:1998vr}), because the size of the box must be much greater than the Planck length. We get the leading contribution, coming from near the horizon,
%%%%%%%%%%%%%%%%%%%%%%%%%%%%%%%%%%%%%%%%%%%%%%%%%%%%%
\begin{eqnarray}
P(\epsilon;\lambda)\approx \frac{2\pi A_{\perp}(\epsilon - \lambda \alpha)^3}{3 \kappa^3 L^2_P}  
\label{es31}
\end{eqnarray}
%%%%%%%%%%%%%%%%%%%%%%%%%%%%%%%%%%%%%%%%%%%%%%%%%%%%
where $A_{\perp}= \int [d^2X] $ is the transverse  area of the side of the box close and tangent to the  horizon. Evidently, the above result holds irrespective of whether we are dealing with the black hole or the cosmological event horizon. 

Considering the quantity $(\epsilon - \lambda \alpha)$ as the total or effective conserved energy $E$,
and taking it to be positive definite (i.e., we are not considering  any negative energy modes or the superradiant instability~\cite{Wald:1984rg}),  the second of Eqs.~(\ref{e15}) now gives
\begin{eqnarray}
Q(\beta) = \frac{4\pi A_{\perp}}{\beta^3 \kappa^3 L^2_P}  
\label{es32}
\end{eqnarray}
We now expand the metric function $-g_{00}\equiv -g_{ab}\chi^a\chi^b=\kappa^2 R^2$ appearing in~(\ref{es16}) around $R=0$ in the radius $L_P$ to get, $-g_{00}=\kappa^2L_p^2$. We can then rewrite $Q(\beta)$ as
\begin{eqnarray}
Q(\beta) = \frac{4\pi A_{\perp} L_P}{(\beta \sqrt{-g_{00}})^3}=   \frac{4\pi L_P A_{\perp}}{\beta^3_{\rm loc} }
\label{es33}
\end{eqnarray}
where $\beta_{\rm loc}= \beta  \sqrt{-g_{00}} $ is the inverse local Tolman temperature, Planck distance away from any of the Killing horizons. The $N$-particle partition function is given by $Q_N=(Q)^N$ and the entropy of the system is given by
$S=(1-\beta_{\rm loc}\partial_{\beta_{\rm loc}})\ln Q_N$,
\begin{eqnarray}
S=N\left[\ln \left(\frac{4\pi L_P A_{\perp} }{N \beta^3_{\rm loc}}\right) +3 \right]
\label{es34}
\end{eqnarray}
which is exactly similar to the static and spherically symmetric spacetimes~\cite{Kolekar:2010py}. This completes our main computational part pertaining the stationary axisymmetric or rotating black holes.

The following point deserves mention.  As evident from Eq.~(\ref{es34}), the entropy associated with normal matter (in this case, an ideal gas) behaves as $\ln (\textrm{Area})$ at the appropriate limit. On the other hand, we know that the black hole entropy scales as $\textrm{Area}$. This shows that there is a large increase  in the number of degrees of freedom associated with a gravitational system before and after the formation of the black hole. That is, as a star collapses to ultimately form a black hole, its entropy just before crossing the would be event horizon scales as logarithm of area, while once the black hole has been formed the entropy scales as area. This  is possibly hinting towards the fact that the gravitational degrees of freedom associated with black holes has to be far more greater in number compared to the degrees of freedom of a normal system. This may have interesting consequences as and when one would seek for a microscopic description of black hole entropy (and some possibilities have been previously discussed in \cite{Padmanabhan:1998jp,Padmanabhan:1998vr}).

The result holds for higher dimensional spacetimes as well, for which the near Killing horizon metric becomes, analogous to Eq.~(\ref{es16}),
\begin{eqnarray}
g_{ab}=-(\kappa R)^{-2}\chi_a\chi_b + R_a R_b + \gamma^{(n-2)}_{ab}
\label{es35}
\end{eqnarray}
where $\gamma^{(n-2)}_{ab}$ is the $(n-2)$-dimensional analogue of the 2-metric $(\mu^{-2}\mu_a\mu_b +f^{-2}\phi_b\phi_b)$, appearing in Eq.~(\ref{es16}). It is then straightforward to obtain after some algebra
\begin{eqnarray}
Q(\beta) = \frac{ (n-1)! \pi^{\frac{n-1}{2}} A_{(n-2), \perp} L_P }  {\Gamma(\frac{n+1}{2}) (n-2) \beta^{n-1}_{\rm loc}}
\label{es36}
\end{eqnarray}
and the entropy
\begin{eqnarray}
S=N\left[ \ln \left( \frac{(n-1)! \pi^{\frac{n-1}{2}} A_{(n-2), \perp} L_P}{\Gamma(\frac{n+1}{2}) (n-2) \beta^{n-1}_{\rm loc} }        \right)  +(n-1) \right]
\label{es37}
\end{eqnarray}
For the sake of completeness, we will next present another derivation of the above result also, by explicitly using the lapse and shift functions. The chief goal is to explicitly demonstrate the existence of the coordinate $R$ above, for a Killing horizon located at some constant radial coordinate say, $r=r_H$. 
%%%%%%%%%%%%%%%%%%%%%%%%%%%%%%%%%%%%%%%%%%%%%%%%%%%%%%%%%%%%%%%%%%%%%%%%%%%%%%%%%%%%%%%%%%%%%%%%%%%
%%%%%%%%%%%%%%%%%%%%%%%%%%%%%%%%%%%%%%%%%%%%%%%%%%%%%%%%%%%%%%%%%%%%%%%%%%%%%%%%%%%%%%%%%%%%%%%%%%%
%%%%%%%%%%%%%%%%%%%%%%%%%%%%%%%%%%%%%%%%%%%%%%%%%%%%%%%%%%%%%%%%%%%%%%%%%%%%%%%%%%%%%%%%%%%%%%%%%%%
\subsection{An alternative derivation of the entropy}
\noindent
Any metric in $3+1$-dimensions, could be expressed in terms of ten independent functions by choosing a particular time foliation, known as the Arnowitt-Deser-Misner decomposition~\cite{Padmanabhan,Arnowitt:1962hi}. However the symmetries of the spacetime for the present case i.e., stationary and axisymmetry demands that five of the off-diagonal metric components must vanish and all the remaining components must be independent of time as well as one of the angular coordinates. Thus any stationary, axisymmetric black hole is characterized by five non-zero functions of $(r,\theta)$, which are denoted by, $N(r,\theta)$, $N^{\phi}(r,\theta)$, $h_{rr}(r\theta)$, $h_{\theta \theta}(r,\theta)$ and $h_{\phi \phi}(r,\theta)$, so that the line element becomes,
%%%%%%%%%%%%%%%%%%%%%%%%%%%%%%%%%%%%%%%%%%%%%%%%%%%%%%%%%%%%%%%%
\begin{align}
ds^{2}&=-N^{2}dt^{2}+h_{rr}dr^{2}+h_{\theta \theta}d\theta ^{2}+h_{\phi \phi}\left(d\phi +N^{\phi}dt\right)^{2}
\nonumber
\\
&=-\left\lbrace N^{2}-h_{\phi \phi}\left(N^{\phi}\right)^{2}\right\rbrace dt^{2}+2h_{\phi \phi}N^{\phi}dtd\phi+h_{rr}dr^{2}+h_{\theta \theta}d\theta ^{2}+h_{\phi \phi}d\phi ^{2}
\end{align}
%%%%%%%%%%%%%%%%%%%%%%%%%%%%%%%%%%%%%%%%%%%%%%%%%%%%%%%%%%%%%%%%%
Among the above functions, the function $N(r,\theta)$ is known as the lapse while $N^{\phi}(r,\theta)$ is known as the shift. The corresponding components for the inverse metric will be,
%%%%%%%%%%%%%%%%%%%%%%%%%%%%%%%%%%%%%%%%%%%%%%%%%%%%%%%%%%%%%%%%%
\begin{align}
g^{tt}=-\frac{1}{N^{2}};\qquad g^{t\phi}=\frac{N^{\phi}}{N^{2}};\qquad g^{rr}=h^{rr}=\frac{1}{h_{rr}};\qquad g^{\theta \theta}=h^{\theta \theta}=\frac{1}{h_{\theta \theta}};\qquad g^{\phi \phi}=\frac{1}{h_{\phi \phi}}-\left(\frac{N^{\phi}}{N}\right)^{2}
\end{align}
%%%%%%%%%%%%%%%%%%%%%%%%%%%%%%%%%%%%%%%%%%%%%%%%%%%%%%%%%%%%%%%%%
such that $h^{\phi \phi}=1/h_{\phi \phi}$. Obviously the above spacetime has two Killing vectors, $\xi^{a}=(\partial_{t})^{a}$ corresponding to time translation and $\phi ^{a}=(\partial_{\phi})^{a}$ corresponding to rotational invariance. The norm of these vectors are 
%%%%%%%%%%%%%%%%%%%%%%%%%%%%%%%%%%%%%%%%%%%%%%%%%%%%%%%%%%%%%%%%
\begin{align}
\xi ^{a}\xi _{a}=g_{tt}&=-\left\lbrace N^{2}-h_{\phi \phi}\left(N^{\phi}\right)^{2}\right\rbrace
\\
\phi ^{a}\phi _{a}=g_{\phi \phi}&=h_{\phi \phi}
\end{align}
%%%%%%%%%%%%%%%%%%%%%%%%%%%%%%%%%%%%%%%%%%%%%%%%%%%%%%%%%%%%%%%%
It is evident from the above equation that the vector $\xi^{a}$ is not timelike everywhere, for the spacetime region in which, $N^{2}<h_{\phi \phi}(N^{\phi})^{2}$, the norm of this vector is positive, indicating existence of the ergosphere. We introduce the following function,
%%%%%%%%%%%%%%%%%%%%%%%%%%%%%%%%%%%%%%%%%%%%%%%%%%%%%%%%%%%%%%%%
\begin{align}
\alpha (r,\theta)\equiv -\frac{\xi ^{a}\phi _{a}}{\phi ^{a}\phi _{a}}=-\frac{g_{t\phi}}{g_{\phi \phi}}=-N^{\phi}
\end{align}
%%%%%%%%%%%%%%%%%%%%%%%%%%%%%%%%%%%%%%%%%%%%%%%%%%%%%%%%%%%%%%%%
which essentially coincides with the shift function, and another vector,
%%%%%%%%%%%%%%%%%%%%%%%%%%%%%%%%%%%%%%%%%%%%%%%%%%%%%%%%%%%%%%%%
\begin{align}
\chi ^{a}=\xi ^{a}-N^{\phi}\phi ^{a}
\end{align}
%%%%%%%%%%%%%%%%%%%%%%%%%%%%%%%%%%%%%%%%%%%%%%%%%%%%%%%%%%%%%%%%
The norm of this vector eventually evaluates to
%%%%%%%%%%%%%%%%%%%%%%%%%%%%%%%%%%%%%%%%%%%%%%%%%%%%%%%%%%%%%%%%%
\begin{align}
\chi ^{a}\chi _{a}&=g_{ab}\chi ^{a}\chi ^{b}=g_{ab}\left(\xi ^{a}-N^{\phi}\phi ^{a}\right)\left(\xi ^{b}-N^{\phi}\phi ^{b}\right)
\nonumber
\\
&=g_{tt}-2g_{t\phi}N^{\phi}+h_{\phi \phi}\left(N^{\phi}\right)^{2}=-N^{2}
\end{align}
%%%%%%%%%%%%%%%%%%%%%%%%%%%%%%%%%%%%%%%%%%%%%%%%%%%%%%%%%%%%%%%%%%
To proceed further, we shall assume that there is some surface, say $\Phi$, on which $N^{2}=0$ and on that surface $\alpha$ i.e., $N^{\phi}$ is a constant (Note that this condition ensures that the geometry depicted above represents a black hole spacetime). Thus on that surface $\chi ^{a}$ is the Killing vector and the Killing horizon of this spacetime corresponds to $\chi^{2}=0$, i.e., the surface $\Phi$. We will also assume that the surface $N^{2}=0$, corresponds to some $r=r_{H}$ and thus one can write $N^{2}$ as,
%%%%%%%%%%%%%%%%%%%%%%%%%%%%%%%%%%%%%%%%%%%%%%%%%%%%%%%%%%%%%%%%%
\begin{align}
N^{2}=\Delta (r)f(r,\theta)
\end{align}
%%%%%%%%%%%%%%%%%%%%%%%%%%%%%%%%%%%%%%%%%%%%%%%%%%%%%%%%%%%%%%%%%%
such that $r=r_{H}$ is a solution of the equation $\Delta(r)=0$. Since $r=r_{H}$ is a Killing horizon, on this surface the norm of the vector $\nabla _{a}r$ must vanish, suggesting, $g^{rr}=0$. Hence we also have,
%%%%%%%%%%%%%%%%%%%%%%%%%%%%%%%%%%%%%%%%%%%%%%%%%%%%%%%%%%%%%%%%%%
\begin{align}
g^{rr}=h^{rr}=\Delta(r)g(r,\theta)
\end{align}
%%%%%%%%%%%%%%%%%%%%%%%%%%%%%%%%%%%%%%%%%%%%%%%%%%%%%%%%%%%%%%%%%%
It is useful to introduce yet another vector field,
%%%%%%%%%%%%%%%%%%%%%%%%%%%%%%%%%%%%%%%%%%%%%%%%%%%%%%%%%%%%%%%%%%%
\begin{align}
R_{a}\equiv \nabla _{a}R \equiv \nabla _{a}\left(\frac{N}{\zeta}\right);\qquad \zeta =\textrm{constant}
\end{align}
%%%%%%%%%%%%%%%%%%%%%%%%%%%%%%%%%%%%%%%%%%%%%%%%%%%%%%%%%%%%%%%%%%
where $\zeta$ is a constant to be determined later. The components of this vector are,
%%%%%%%%%%%%%%%%%%%%%%%%%%%%%%%%%%%%%%%%%%%%%%%%%%%%%%%%%%%%%%%%%%
\begin{align}
R_{r}&=\frac{1}{\zeta}\partial _{r}\left(\sqrt{\Delta}\sqrt{f}\right)=\frac{\sqrt{f}\Delta '}{2\zeta \sqrt{\Delta}}
+\frac{\sqrt{\Delta}f'}{2\zeta \sqrt{f}};\qquad R^{r}=h^{rr}R_{r}=\frac{g\sqrt{f}\Delta '\sqrt{\Delta}}{2\zeta}
+\frac{\Delta ^{3/2}gf'}{2\zeta \sqrt{f}}
\\
R_{\theta}&=\frac{\sqrt{\Delta}\partial _{\theta}f}{2\zeta \sqrt{f}}
\end{align}
%%%%%%%%%%%%%%%%%%%%%%%%%%%%%%%%%%%%%%%%%%%%%%%%%%%%%%%%%%%%%%%%%%
where `prime' denotes derivative with respect to the radial coordinate $r$. For stationary spacetime one can prove that  $(\nabla _{a}N)(\nabla ^{a}N)$ is a constant in the limit $r\rightarrow r_{H}$, defined as the square of the surface gravity $\kappa$, leading to
%%%%%%%%%%%%%%%%%%%%%%%%%%%%%%%%%%%%%%%%%%%%%%%%%%%%%%%%%%%%%%%%%%%
\begin{align}
\lim _{r\rightarrow r_{H}}R_{a}R^{a}&=\frac{\kappa ^{2}}{\zeta ^{2}}
=g^{rr}\left(\frac{\sqrt{f}\Delta '}{2\zeta \sqrt{\Delta}}\right)^{2}=\frac{gf\Delta '^{2}}{4\zeta ^{2}}
\end{align}
%%%%%%%%%%%%%%%%%%%%%%%%%%%%%%%%%%%%%%%%%%%%%%%%%%%%%%%%%%%%%%%%%%%
Requiring $R_{a}R^{a}=1$ on the surface $r=r_{H}$, we  obtain,
%%%%%%%%%%%%%%%%%%%%%%%%%%%%%%%%%%%%%%%%%%%%%%%%%%%%%%%%%%%%%%%%%%%
\begin{align}
\zeta =\kappa =\sqrt{\frac{gf\Delta'^{2}}{4}}\Bigg\vert _{r=r_{H}}
\end{align}
%%%%%%%%%%%%%%%%%%%%%%%%%%%%%%%%%%%%%%%%%%%%%%%%%%%%%%%%%%%%%%%%%%%%
Since this has to be constant one obtains that $f(r,\theta)g(r,\theta)$ should be independent of $\theta$ on the Killing horizon. Thus we get, $N^{2}=\kappa ^{2}R^{2}$. Using this expression for the lapse function $N$, the metric can be written as,
%%%%%%%%%%%%%%%%%%%%%%%%%%%%%%%%%%%%%%%%%%%%%%%%%%%%%%%%%%%%%%%%%%%%
\begin{align}
g^{ab}=-\frac{1}{\kappa ^{2}R^{2}}\chi ^{a}\chi ^{b}+h^{ab}
\end{align}
%%%%%%%%%%%%%%%%%%%%%%%%%%%%%%%%%%%%%%%%%%%%%%%%%%%%%%%%%%%%%%%%%%%%
where of course, $h^{ab}$ is purely spatial. To rewrite it as fit for the the near horizon limit, we note that the spatial metric can be further expended leading to,
%%%%%%%%%%%%%%%%%%%%%%%%%%%%%%%%%%%%%%%%%%%%%%%%%%%%%%%%%%%%%%%%%%%%
\begin{align}
h^{ab}&=h^{\phi \phi}\phi ^{a}\phi ^{a}+h^{\theta \theta}(\partial _{\theta})^{a}(\partial _{\theta})^{b}+h^{rr}(\partial _{r})^{a}(\partial _{r})^{b}
\nonumber
\\
&\stackrel{r\simeq r_{H}}{=}h^{\phi \phi}\phi ^{a}\phi ^{a}+h^{\theta \theta}(\partial _{\theta})^{a}(\partial _{\theta})^{b}+R^{a}R^{b}+\mathcal{O}(\Delta)
\end{align}
%%%%%%%%%%%%%%%%%%%%%%%%%%%%%%%%%%%%%%%%%%%%%%%%%%%%%%%%%%%%%%%%%%%%%
The last line follows from the fact that, $(R^{r})^{2}=\Delta (r) g(r,\theta) +\mathcal{O}(\Delta ^{2})$, coinciding with the correct leading order behavior for $h^{rr}$, while both $R^{r}R^{\theta}$ and $R^{\theta}R^{\theta}$ are $\sim \mathcal{O}(\Delta)$ and do not contribute in the near horizon limit. Thus the full metric in the near horizon regime becomes Rindler-like,
%%%%%%%%%%%%%%%%%%%%%%%%%%%%%%%%%%%%%%%%%%%%%%%%%%%%%%%%%%%%%%%%%%%%%%
\begin{align}
g^{ab}\stackrel{r\simeq r_{H}}{=}h^{\phi \phi}\phi ^{a}\phi ^{a}+h^{\theta \theta}(\partial _{\theta})^{a}(\partial _{\theta})^{b}+R^{a}R^{b}-\frac{1}{\kappa ^{2}R^{2}}\chi ^{a}\chi ^{b}
\end{align}
%%%%%%%%%%%%%%%%%%%%%%%%%%%%%%%%%%%%%%%%%%%%%%%%%%%%%%%%%%%%%%%%%%%%%%%
Thus given any stationary and axisymmetric black hole spacetime, the metric near the black hole horizon can always be written in a Rindler form. Since we are chiefly interested in the behavior of a box full of ideal gas in the near horizon regime, the above Rindler form straightforwardly suggests that the entropy of the ideal gas should scale as area. To see this explicitly, we will first compute the probability of a particle in the ideal gas to have an energy $E$. Since the probability distribution involves an integral over phase space (see Eq.~(\ref{e16})) which extends upto the energy $E$ in momentum space, one can integrate out the momentum degrees of freedom and finally obtains,
%%%%%%%%%%%%%%%%%%%%%%%%%%%%%%%%%%%%%%%%%%%%%%%%%%%%%%%%%%%%%%%%%%%%%%
\begin{align}
P(E)=\frac{4\pi}{3}\int d^{3}x\sqrt{h}\left(\frac{E^{2}}{N^{2}}-m^{2}\right)^{3/2}
\end{align}
%%%%%%%%%%%%%%%%%%%%%%%%%%%%%%%%%%%%%%%%%%%%%%%%%%%%%%%%%%%%%%%%%%%%%%
where $E=-p_{a}\xi ^{a}$, linearly dependent on the conserved energy and angular momentum of the particle and we may identify $N\equiv \tilde{\beta}$ in our earlier analysis. Note that the above result holds for any stationary and axisymmetric spacetimes. Finally specializing to the Kerr metric, the integral leading to the probability distribution near the horizon becomes,
%%%%%%%%%%%%%%%%%%%%%%%%%%%%%%%%%%%%%%%%%%%%%%%%%%%%%%%%%%%%%%%%%%%%%%%
\begin{align}
P(E)=\frac{8\pi ^{2}E^{3}}{3}\int dr d\theta ~\frac{\Sigma ^{4}\sin \theta}{\rho ^{2}\Delta ^{2}}
\end{align}
%%%%%%%%%%%%%%%%%%%%%%%%%%%%%%%%%%%%%%%%%%%%%%%%%%%%%%%%%%%%%%%%%%%%%%%%
where, $\Delta =r^{2}+a^{2}-2Mr$, $\rho ^{2}=r^{2}+a^{2}\cos ^{2}\theta$ and $\Sigma =(r^{2}+a^{2})^{2}-a^{2}\Delta \sin ^{2}\theta$. In order to simplify the above result we perform a change of coordinates, $(r,\theta)\rightarrow (R,\theta)$, where $R=N/\kappa=\rho \sqrt{\Delta}/(\kappa \Sigma)$, with the last equality holding for the Kerr metric. The Jacobian associated with the above coordinate transformation becomes
%%%%%%%%%%%%%%%%%%%%%%%%%%%%%%%%%%%%%%%%%%%%%%%%%%%%%%%%%%%%%%%%%%%%%%%%%
\begin{align}
J=\textrm{det.}
\left(\begin{array}{ll}
       \frac{1}{\sqrt{g\Delta}} & \frac{\sqrt{\Delta}\partial _{\theta}f}{2\kappa \sqrt{f}}\\
       -\frac{2\kappa \sqrt{f}}{\Delta \sqrt{g}\partial _{\theta}f} &~~~~~ 1\\
      \end{array}\right)
=\frac{2}{\sqrt{g\Delta}}=\frac{2\rho}{\sqrt{\Delta}}
\end{align}
%%%%%%%%%%%%%%%%%%%%%%%%%%%%%%%%%%%%%%%%%%%%%%%%%%%%%%%%%%%%%%%%%%%%%%%%%%
The Jacobian evaluated above is completely general and applies to any stationary and axisymmetric spacetimes, however for illustration purpose in the last expression we have used the results for the Kerr metric. Then the integral leading to probability distribution can be rewritten in the new coordinate system $(R,\theta)$ as,
%%%%%%%%%%%%%%%%%%%%%%%%%%%%%%%%%%%%%%%%%%%%%%%%%%%%%%%%%%%%%%%%%%%%%%%%%
\begin{align}
P(E)&=\frac{8\pi ^{2}E^{3}}{3}\int dR d\theta ~J^{-1}\frac{\Sigma ^{4}\sin \theta}{\rho ^{2}\Delta ^{2}}
=\frac{4\pi ^{2}E^{3}}{3}\int dR d\theta ~\frac{\Sigma ^{4}\sin \theta}{\rho ^{3}\Delta ^{3/2}}
\nonumber
\\
&=\frac{4\pi ^{2}E^{3}\Sigma}{3\kappa ^{3}}\int dR d\theta ~\frac{\sin \theta}{R^{3}}=\frac{2\pi E^{3}A_{\perp}}{3\kappa ^{3}}\int \frac{dR}{R^{3}}
=\frac{2\pi E^{3}A_{\perp}}{3\kappa ^{3}L_{P}^{2}}
\end{align}
%%%%%%%%%%%%%%%%%%%%%%%%%%%%%%%%%%%%%%%%%%%%%%%%%%%%%%%%%%%%%%%%%%%%%%%
where in the last line we have used the fact that the box is Planck length away from the black hole. Given this probability distribution, one immediately observes that the distribution itself scales with area and it is the source of the area dependence of entropy for the ideal gas. Since the distribution is known to us, one can compute the partition function using Eq.~(\ref{e15}), where $\beta$ will be the inverse temperature associated with the black hole. Thus we arrive at the following expression for the partition function,
%%%%%%%%%%%%%%%%%%%%%%%%%%%%%%%%%%%%%%%%%%%%%%%%%%%%%%%%%%%%%%%%%%%%%%%
\begin{align}
Q(\beta)=\frac{2\pi A_{\perp}}{\kappa ^{3}L_{P}^{2}}\int dE ~E^{2}e^{-\beta E}=\frac{2\pi A_{\perp}}{\kappa ^{3}L_{P}^{2}}\frac{\partial ^{2}}{\partial \beta ^{2}}\left(\frac{1}{\beta}\right)=\frac{4\pi A_{\perp}}{\beta ^{3}\kappa ^{3}L_{P}^{2}}
\end{align}
%%%%%%%%%%%%%%%%%%%%%%%%%%%%%%%%%%%%%%%%%%%%%%%%%%%%%%%%%%%%%%%%%%%%%%%
Introducing a redshifted local temperature, $\beta _{\rm loc}=N\beta=\beta (\kappa L_{P})$, where $N=\kappa L_{P}$ is the redshift factor, we finally obtain
%%%%%%%%%%%%%%%%%%%%%%%%%%%%%%%%%%%%%%%%%%%%%%%%%%%%%%%%%%%%%%%%%%%%%%%
\begin{align}
Q(\beta_{\rm loc})=\frac{4\pi A_{\perp}}{\beta _{\rm loc}^{3}}
\end{align}
%%%%%%%%%%%%%%%%%%%%%%%%%%%%%%%%%%%%%%%%%%%%%%%%%%%%%%%%%%%%%%%%%%%%%%%%%%
which matches exactly with our earlier result. Thus starting from the lapse, shift functions and imposing the symmetries, one ends up with the Rindler form of the metric in the near horizon region and an area scaling for entropy. This provides another way of looking into the problem by explicitly using coordinates and perhaps is interesting from the gravitational dynamics standpoint.  
%%%%%%%%%%%%%%%%%%%%%%%%%%%%%%%%%%%%%%%%%%%%%%%%%%%%%%%%%%%%%%%%%%%%%%%%%%%%%%%%%%%%%%%%%%%%%%%%%%%
%%%%%%%%%%%%%%%%%%%%%%%%%%%%%%%%%%%%%%%%%%%%%%%%%%%%%%%%%%%%%%%%%%%%%%%%%%%%%%%%%%%%%%%%%%%%%%%%%%%
%%%%%%%%%%%%%%%%%%%%%%%%%%%%%%%%%%%%%%%%%%%%%%%%%%%%%%%%%%%%%%%%%%%%%%%%%%%%%%%%%%%%%%%%%%%%%%%%%%%
\section{The cosmological spacetimes}

%%%%%%%%%%%%%%%
\subsection{The metric and suitable coordinate choice } 
%%%%%%%%%%%%%%%%%
\noindent In this section we shall further apply the formalism for cosmological spacetimes which are essentially non-stationary. Let us start with the Friedmann-Lemaitre-Robertson-Walker spacetime with flat spatial sections,
\begin{eqnarray}
ds^2=-dt^2+ a^2(t)\left[dx^2+dy^2+dz^2\right]
\label{e1}
\end{eqnarray}
The first of the Einstein equations reads,
\begin{eqnarray}
H^2(t)=\left(\frac{\dot {a}}{a}\right)^2 = 8\pi  \rho(t) + \Lambda 
\label{e2}
\end{eqnarray}
For any generic matter field with a linear equation of state, $p(t)=w \rho(t)$, where $w$ is a constant, the energy-momentum conservation corresponding to Eq.~(\ref{e1}) gives
$$\rho(t)=\frac{\rho_0}{a^{3(1+w)}(t)}$$ 
where $\rho_0$ is a constant. Plugging this into Eq.~(\ref{e2}), we get
\begin{eqnarray}
a(t) = \left(\frac{8\pi \rho_0}{\Lambda} \right)^{\frac{1}{3(1+w)}} \left(\sinh \frac{3(1+w)H_0 t}{2}\right)^{\frac{2}{3(1+w)}} 
\label{e3c}
\end{eqnarray}
where $H_0=\sqrt{\Lambda/3}$, as earlier. We shall in particular be interested in cold dark matter ($w=0$) and radiation ($w=1/3$). For the first, the scale factor reads,
\begin{eqnarray}
a(t)=  \left(\frac{8\pi \rho_0}{\Lambda} \right)^{\frac13} \sinh^{2/3} \frac{3H_0 t}{2}  
\label{e3}
\end{eqnarray}
For $t \to \infty$, the above scale factor evolves to that of the de Sitter, $a(t)\sim e^{H_0 t} $. Likewise for the radiation we have, 
\begin{eqnarray}
a(t)= \left(\frac{8\pi \rho_0}{\Lambda} \right)^{\frac14}\sinh^{1/2} 2H_0 t.
\label{e4}
\end{eqnarray}
Rewriting Eq.~(\ref{e1}) in terms of spherical polar coordinates $(R,~\theta,~\phi)$ and introducing the proper or the physical radius $r=R a(t)$, we get
\begin{eqnarray}
ds^2= -(1-H^2(t) r^2)dt^2-2H(t)rdrdt +dr^2 +r^2 d\Omega^2
\label{e5}
\end{eqnarray}
The Hubble horizon radius, $H^{-1}(t)$, is manifest in the above coordinate system. The coordinate singularity at the surface $r=H^{-1}(t)$ is not in general a null surface for any scale factor like $a(t)=a_0 \sinh^p \frac{H_0 t}{p}$, as can be seen by computing the norm of the 1-form $\nabla _a (r H(t))$ normal to that surface,
\begin{eqnarray}
g^{ab}\nabla _a (r H(t))\nabla _b (r H(t)) = -\frac{H_0^2/p}{ \sinh^2 (H_0t/p)}\left[\frac{1}{p\cosh^2 (H_0t /p)}-2 \right] \neq 0
\label{e5'}
\end{eqnarray}
while the equality certainly holds in the de Sitter limit $ H_0 t \to \infty $.

Let us suppose that we are dealing with the empty de Sitter space, so that $H(t)$ in Eq.~(\ref{e5}) coincides with $H_0$. One can then introduce a new
timelike coordinate, $T$ (see, e.g.~\cite{Kastor:1992nn}), 
\begin{eqnarray}
t= T+ \frac{1}{2H_0} \ln \vert 1- H_0^2 r^2\vert
\label{e5''}
\end{eqnarray}
to get
\begin{eqnarray}
ds^2= -(1-H_0^2r^2)dT^2 +(1-H_0^2r^2)^{-1}dr^2 +r^2 d\Omega^2,
\label{e6}
\end{eqnarray}
which is manifestly static inside the cosmological event horizon radius, $H_0^{-1}$. However, since we are interested in a time dependent Hubble expansion rate, we cannot define a coordinate transformation like the above by just replacing $H_0$ with $H(t)$ \footnote{Note that while dealing with stationary spacetimes in Sec.~II, we used `$t$' instead of `$T$' for the timelike coordinate. We use `$T$' here just because we wish to retain `$t$' as the cosmological time.}.

On the other hand, using indeed the transformation of Eq.~(\ref{e5''}), the metric of Eq.~(\ref{e5}) can be rewritten as
\begin{eqnarray}
&&ds^2= -\left(1-r^2H^2(T,r)\right)dT^2 +2r\left( \frac{H_0 \left(1-H^2(T,r) r^2\right)}{1-H_0^2 r^2 } - H(T,r)\right) dT dr  \nonumber\\ && +\left(1+ \frac{2r^2 H(T,r) H_0 }{1-  H_0^2 r^2  } - \frac{H_0^2 r^2 \left(1-r^2 H^2(T,r)\right)}{\left(1- H_0^2 r^2\right)^2 }  \right)dr^2  +r^2 d\Omega^2
\label{e8}
\end{eqnarray}
Clearly the current Hubble radius, $H^{-1}(t)$, is smaller than the cosmological horizon radius $H_0^{-1}$, of the de Sitter space. Accordingly, it is easy to see that the second term in $g_{rr}$ is always greater than that of the third term which is negative. Also, in the limit $H(T,r) \to H_0$, we recover the static patch of the de Sitter space. Putting these all in together, it is clear that the coordinate system used in Eq.~(\ref{e8}) is well defined within the current Hubble radius. The above coordinate system would be much useful for our purpose of analyzing systems asymptotically evolving to the de Sitter. 

The metric in Eq.~(\ref{e8}) can also be generalized easily to include a central mass/black hole. The version of Eq.~(\ref{e1}) is now given by the McVittie spacetime written in the spherical polar coordinates~\cite{Gao:2004wn, Kaloper:2010ec},
\begin{eqnarray}
&&ds^2= - \left(    \frac{1-\frac{M}{2Ra(t)} } {1+\frac{M}{2Ra(t)}}   \right)^2 dt^2 +a^2(t) \left( 1+ \frac{M}{2Ra(t)} \right)^4 \left(dR^2 +R^2 d\Omega^2\right)
\label{e9}
\end{eqnarray}
Defining the proper radius $r=Ra(t)\left( 1+ M/2Ra(t) \right)^2  $, the above metric can be rewritten as 
\begin{eqnarray}
&&ds^2= - \left(1-\frac{2M}{r}-H^2r^2 \right) dt^2 -2Hr\left(1-\frac{2M}{r}\right)^{-\frac12} dr dt +\left(1-\frac{2M}{r}\right)^{-1}dr^2   +r^2 d\Omega^2.
\label{e10}
\end{eqnarray}
For $H=H_0$ the above simply represents the Schwarzschild-de Sitter spacetime. In particular, defining $$t= T - H_0 \int \frac{rdr}{(1-2M/r)^{\frac12} (1-2M/r -H_0^2 r^2) },$$ the above metric with $H=H_0$ takes the standard form of the Schwarzschild-de Sitter,
\begin{eqnarray}
&&ds^2= - \left(1-\frac{2M}{r}-H_0^2r^2 \right) dT^2  +\left(1-\frac{2M}{r}-H_0^2 r^2\right)^{-1}dr^2   +r^2 d\Omega^2
\label{e10n}
\end{eqnarray}
whereas for a generic scale factor, we find, in place of Eq.~(\ref{e8}) or Eq.~(\ref{e10}) 
\begin{eqnarray}
&&ds^2= - \left(1-\frac{2M}{r}-H^2(T,r)r^2 \right) dT^2  +   \frac{2rdT dr}{\sqrt{(1-2M/r)}} \left(\frac{H_0  \left(1-\frac{2M}{r}-H^2(T,r)r^2 \right)   }{ \left(1-\frac{2M}{r}-H_0^2r^2 \right) }-H(T,r)\right) \nonumber\\
&&+        \frac{dr^2 }{(1-2M/r)} \left(1+ \frac{2H(T,r) H_0 r^2 }{\left(1-\frac{2M}{r}-H_0^2r^2 \right)} -\frac{H_0^2r^2 \left(1-\frac{2M}{r}-H^2(T,r)r^2 \right)}{\left(1-\frac{2M}{r}-H_0^2r^2 \right)^2} \right)  +r^2 d\Omega^2
\label{e11}
\end{eqnarray}
in which, setting $H\to H_0$ recovers the Schwarzschild-de Sitter spacetime. Note that the above metric could also represent the dynamical horizon of a black hole. The dynamical cosmological and black hole apparent horizons are given respectively by the larger and the smaller roots of $g_{TT}=0$. 

First, we shall be interested in the dynamics of spacetimes in Eq.~(\ref{e8}) or Eq.~(\ref{e11}) when they are close to the de Sitter or Schwarzschild-de Sitter. In that case the Hubble rate, $H=\dot{a}/a$, corresponding to the scale  factors of Eqs.~(\ref{e3}),~(\ref{e4}) take the form $H_0\left(1+\delta H(T,r)\right)$ (with $H_0t \gg1$), where
\begin{eqnarray}
&&\delta H(T,r)\approx 6H_0 t(T,r) e^{-3H_0 t(T,r)} \qquad ({\rm for}~\Lambda{\rm CDM})\nonumber\\
&&\delta H(T,r)\approx 8H_0 t(T,r) e^{-4H_0 t(T,r)} \qquad ({\rm for}~\Lambda~{\rm radiation})
\label{e12}
\end{eqnarray}
The above relations show that the $\Lambda$-radiation universe evolves faster to the de Sitter compared to $\Lambda{\rm CDM}$, just because of the fact that the pressure of the radiation dilutes itself faster than the cold dark matter.
Eq.~(\ref{e8}) takes the form up to ${\cal O}(\delta H)$,
\begin{eqnarray}
&&ds^2= -\left(1-r^2H_0^2 -2r^2H_0^2\delta H(T,r)\right)dT^2 - 2\delta H(T,r) H_0 r\frac{1+H_0^2r^2}{1-H_0^2r^2} dT dr  \nonumber\\&&+\left(1- H_0^2 r^2\right)^{-1} \left( 1+\frac{2\delta H(T,r) r^2 H_0^2}{1- H_0^2 r^2 } \right)dr^2  +r^2 d\Omega^2
\label{e13}
\end{eqnarray}
We note that we  must  take the $\delta H \to 0$ limit before taking the limit $r \to H_0^{-1}$, as the de Sitter cosmological  horizon radius is larger than the dynamical Hubble radius $H^{-1}(T,r)$, and our coordinate system is well behaved only up to the latter. Likewise, Eq.~(\ref{e11}) becomes
\begin{eqnarray}
&&ds^2= - \left(1-\frac{2M}{r}-r^2H_0^2 -2r^2H_0^2\delta H(T,r) \right) dT^2  -   \frac{2H_0 \delta H(T,r)r}{\sqrt{(1-2M/r)}} \frac{ \left(1-\frac{2M}{r}+H_0^2r^2 \right) }{ \left(1-\frac{2M}{r}-H_0^2r^2 \right) }dT dr \nonumber\\
&&+        \frac{dr^2 }{(1-2M/r-H_0^2r^2)} \left(1+ \frac{2 H_0^2 r^2 \delta H(T,r)}{(1-2M/r-H_0^2r^2)}   \right)   +r^2 d\Omega^2
\label{e14}
\end{eqnarray}
Spacetimes in Eqs.~(\ref{e13}), (\ref{e14}), now being close to  the de Sitter or the Schwarzschild-de Sitter (respectively) can  be treated as quasi-static. We shall use these two forms of the actual metrics in order to explicitly evaluate the entropy
of a box filled with ideal gas, in the next section.
%%%%%%%%%%%%%%%
\subsection{Calculation of the entropy } 
%%%%%%%%%%%%%%%%%
\noindent
Let us consider Eq.~(\ref{e13}), in which we need to first obtain a suitable spatial hypersurface and an orthogonal timelike vector field, $\chi^a$. Certainly, the most convenient choice of the spatial hypersurface would be the $(r,\theta,\phi)$ hypersurfaces. Since $(\partial_T)^a$ is not orthogonal to that hypersurface, we define a vector field 
\begin{eqnarray}
\chi^a=(\partial_T)^a- \frac{(\partial_T\cdot \partial_r)}{(\partial_r\cdot \partial_r)}(\partial_r)^a
\label{e17}
\end{eqnarray}
Clearly, the above construction is analogous to what we did in the stationary axisymmetric case with the difference that $(\partial_r)^a$ is not a Killing vector field here. We have the  norm of  $\chi^a$
\begin{eqnarray}
\chi^a\chi_a = -\tilde{\beta}^2=-\left(1-r^2H_0^2 -2r^2H_0^2\delta H(T,r)\right)+{\cal O} (\delta H)^2,
\label{e18}
\end{eqnarray}
guaranteeing that $\chi^a$ is indeed timelike within the current Hubble radius. Since by construction $\chi_a(\partial_r)^a=0$, we shall use $\chi^a$ in Eq.~(\ref{e16}). 
Putting these all in together, and using the momentum dispersion relation, $p_ap^a=-m^2$, the phase space volume, Eq.~(\ref{e16}), becomes
\begin{eqnarray}
P(E(T))=\frac{4\pi}{3}\int dr d\theta\sin\theta  d\phi \frac{ r^2}{\sqrt{1- H_0^2 r^2}}\left(1+ \frac{r^2 H_0^2 \delta H(T,r)}{1- H_0^2 r^2}\right)  \left( \frac{E^2(r,T)}{1- H_0^2 r^2-2r^2H_0^2 \delta H(T,r) }-m^2 \right)^{\frac32} 
\label{e19} 
\end{eqnarray}
where we have defined $E(r,T)=-p_a\chi^a$ and since $\chi^a$ is not a Killing vector field here, this quantity is not a constant.
Putting $H_0=0$ corresponds to the  flat spacetime, whereas putting $\delta H(T,r)=0$  to the de Sitter spacetime, which we shall briefly consider first, following~\cite{Kolekar:2010py}. Note that in this limit the vector field $\chi^a$ is  Killing and and $E$ should be regarded as the conserved energy.  

Once again, we need to impose the Planck scale cut off in probing the near horizon geometry. But we can now  use the radial coordinate itself for this purpose, unlike in the stationary axisymmetric spacetimes. Let us imagine that the box is close to the cosmological event horizon, with its sides located at $r_a$ (outer side, closer to the horizon) and $r_b$ (the inner side). Let $r_a =H_0^{-1}- L_a$ and $r_b= H_0^{-1}-L_b$.  We choose $L_a$ to be such that it is related to the Planck length $L_P$ as  
\begin{eqnarray}
L_P =\int_{H_0^{-1}-L_a}^{H_0^{-1}} \frac{dr}{\sqrt{1-H_0^2 r^2}}=\sqrt{\frac{2L_a}{H_0}}
\label{e20} 
\end{eqnarray}
We assume that $L_P\ll H_0^{-1}$ which is very natural owing to the observed tiny value of the cosmological constant. The cosmological event horizon has a negative surface gravity $-\kappa_C=-H_0$, owing to the repulsive effect. The integral in Eq.~(\ref{e19}) for the de Sitter space becomes, as $r\to H_0^{-1}$,  
\begin{eqnarray}
P(E)=\frac{4\pi E^3}{3} \int_{r_b}^{r_a} dr \sin\theta d\theta d\phi \frac{r^2}{\left(1- H_0^2 r^2\right)^2}   
\label{e21} 
\end{eqnarray}
which yields, after using Eq.~(\ref{e20}) at the leading order
\begin{eqnarray}
P(E)\approx\frac{2\pi E^3}{3}\frac{A_{\perp}}{L_P^2 \kappa_C^3}   
\label{e22} 
\end{eqnarray}
where as earlier  $A_{\perp}$ is the transverse area of the box, tangent to the cosmological event horizon. The partition function is obtained from the second relation in Eq.~(\ref{e15}),
\begin{eqnarray}
Q(\beta)= \frac{4\pi A_{\perp}   }{\beta^3 \kappa_C^3 L_P^2}
\label{e23} 
\end{eqnarray}
Now, using Eq.~(\ref{e20}) we find $(1-H_0^2r^2)\vert_{r=r_a} \approx L_P^2 H_0^2$. Putting this back into the above equation, we get
\begin{eqnarray}
Q(\beta)=  \frac{4\pi A_{\perp} L_P  }{\beta^3_{\rm loc}(r_a)}
\label{e24} 
\end{eqnarray}
where $\beta_{\rm loc}(r_a)= \beta \sqrt{L_P^2 H_0^2}$ is the inverse of the blue shifted local temperature as earlier. For $N$  non-interacting particles, we get the  partition function to be $Q_N=Q^{N}(\beta)$ and the entropy,  
\begin{eqnarray}
S= N\left[\ln \left(\frac{4\pi L_P A_{\perp} }{N\beta^3_{\rm loc}(r_a) }\right) +3\right]
\label{e25} 
\end{eqnarray}
which, as expected, is exactly formally similar to the result for stationary axisymmetric  spacetimes, Eq.~(\ref{es34}).

Let us consider the  generic case now. We cannot evaluate Eq.~(\ref{e19}) along the above lines as we do not know the behaviour of the function $E(r,T)$. However, we can still determine the scaling properties of the  entropy when the spacetime is close to the de Sitter, as follows.

The Hubble horizon radius $r_C(T)$, is given in this case by $g_{TT}=-(1- H_0^2 r^2-2r^2H_0^2 \delta H)=0$, 
\begin{eqnarray}
r_C(T)\approx \frac{1}{H_0}(1-\delta H)
\label{e28n'} 
\end{eqnarray}
When one side of the box is close to the Hubble horizon, the integral of Eq.~(\ref{e19}) diverges. In order to regularize this integral, we need to define a time-dependent cut off, $L_P(T)$ as,
\begin{eqnarray}
L_P(T)=\int_{r_C(T)-L_a}^{r_C(T)}\frac{dr}{\sqrt{1-H_0^2r^2-2r^2H_0^2 \delta H}}
\label{e28p} 
\end{eqnarray}
where, as in Eq.~({\ref{e20}), $(r_C(T)-L_a)$ is the (time-dependent) radial coordinate of the side of the  box which is close to the Hubble horizon. (If we set  $\delta H =0$ above, we recover the Planck length $L_P$, as the cut off). Assuming that the length scale of the box is much smaller compared to the horizon length scale, we have
\begin{eqnarray}
L_P(T) \approx \sqrt{\frac{2L_a}{H}}	
\label{e29}
\end{eqnarray}
We can also define the `surface gravity' of the Hubble horizon as, $\kappa^2(T):=\frac{1}{4\tilde{\beta}^2} g^{ab}(\nabla_a \tilde{\beta}^2) (\nabla_b \tilde{\beta}^2)$, which becomes, to the leading order,
\begin{eqnarray}
-\kappa(T)\approx -H(T)=-H_0(1+\delta H)+{\cal O}(\delta H)^2
\label{e26n} 
\end{eqnarray}
Now, by noting that  in the leading order we have 
$$\sqrt{1-H_0^2 r^2}=\sqrt{1-r^2 H^2(T,r)} \left[1+\frac{H_0^2 r^2 \delta H}{1-H_0^2 r^2} \right],$$
Eq.~(\ref{e19}) becomes, when the box is close to the Hubble horizon,
\begin{eqnarray}
P(E(T))\approx\frac{4\pi}{3} \int_{r_b}^{r_a} dr \sin\theta d\theta d\phi \frac{r^2 E^3(r,T)}{  \left(1-r^2H^2(T,r)\right)^{2}}
\label{e26} 
\end{eqnarray}
which is formally similar to the Killing horizon case, certainly due to the quasi-de Sitter approximation.
Since in this limit $\delta H \ll 1$ in Eqs.~(\ref{e12}),
it follows that the function $E(r,T)$ varies slowly compared to the divergent denominator of the above integrand. Also,  we can pull out the $r^2=r^2_{C}(T)$ term outside the integration. Putting  all these in together and performing the  angular integration, we obtain
\begin{eqnarray}
P(E(T))\approx\frac{4\pi E^3(T) A_{\perp}(T)}{3}\int_{r_b=r_C(T)-L_b}^{r_a=r_C(T)-La} \frac{dr}{ \left(1-r^2H^2(T,r)\right)^{2}}
\label{e30'} 
\end{eqnarray}
where, $A_{\perp}(t)=\Omega\, r^2_C(T)$ is the section of the transverse area of the Hubble horizon, parallel to the approaching side of the box, with $\Omega$ being the relevant solid angle. This expression could further be simplified as 
\begin{eqnarray}
P(E(T))\approx \frac{\pi E^3(T) A_{\perp}(T)}{3 }\int_{r_b=r_C(T)-L_b}^{r_a=r_C(T)-L_a} \frac{dr}{ \kappa^2(T) \left(r_C(T)-r\right)^{2}}
\label{e28} 
\end{eqnarray}
and could easily be evaluated by using Eqs.~(\ref{e28n'}),~(\ref{e29}),~ (\ref{e26n}) and once again the fact that $\delta H$ is a slowly varying quantity:
\begin{eqnarray}
P(E(T))\approx \frac{2\pi E^3(T) A_{\perp}(T)}{3 \kappa^3(T) L^2_P(T) }
\label{e29'} 
\end{eqnarray}
The corresponding expression for the entropy becomes, 
\begin{eqnarray}
S\approx  N\left[\ln \left(\frac{4\pi L_P(T) A_{\perp}(T) }{N\beta^3_{\rm loc} (T)}\right) +3 \right]
\label{e30} 
\end{eqnarray}
which is formally similar to the result in the case of the de Sitter spacetime. Since all the time dependent quantities appearing  above have smooth de Sitter limit, the above expression for the entropy asymptotes to Eq.~(\ref{e25}) at late times. (To the best of our knowledge, this is the first demonstration of the emergence of the entropy-area relation in a physical scenario for a Hubble horizon.) Even though we have performed the above computations in the asymptotic de Sitter limit, we shall argue at the end of this section that such area scaling should occur in the case of a generic Hubble horizon as well. In other words, our analysis shows that the area scaling for the entropy for a  Killing horizon, Eq.~(\ref{e25}), should come from a smooth limit from a non-killing or dynamical horizon.

There is, however, one caveat to the above derivation. The notion of the entropy of a given system is usually related to the lack of information to the observer, as happens in the case of a null surface e.g.,~\cite{Chakraborty:2016dwb}. For such surfaces, we get an area scaling due to the piling up of accessible states near the horizon~\cite{Padmanabhan:1998jp, Padmanabhan:1998vr}, for the observers who do not cross the surface. This also makes the Planck scale cut off imposed earlier to be physically meaningful. On the other hand, since the Hubble horizons are not null surfaces, the time dependent cut off $L_P(T)$, used in Eq.~(\ref{e29'})  should perhaps be regarded as merely a regulator to handle the divergence of the near-horizon integral. 

The above analysis works for  the McVittie spacetime of Eq.~(\ref{e14}) as well, for which the dynamical black hole $(r_H(T))$ and the Hubble horizons ($r_C(T)$) are the two real positive roots of  
$$\left(1-2M/r-r^2H_0^2 -2r^2H_0^2\delta H(T,r) \right)=0$$
The integral of Eq.~(\ref{e26}) now modifies to, when close to either of these horizons,
\begin{eqnarray}
P(E(T))\approx\frac{4\pi}{3} \int_{r_b}^{r_a} dr \sin\theta d\theta d\phi \frac{r^2 E^3(T)}{  \left(1-2M/r-r^2H_0^2 -2r^2H_0^2\delta H\right)^{2}}
\label{e31} 
\end{eqnarray}
By taking one side of the box close to either of the horizons, the above integral could be evaluated as earlier to obtain similar results, by expanding various terms to ${\cal O}(\delta H)$ from their stationary (i.e. Schwarzschild-de Sitter) values.

Let us finally consider the generic case of Eq.~(\ref{e8}) and take the near Hubble horizon limit, 
\begin{eqnarray}
P(E(T))\approx\frac{4\pi}{3} \int_{r_b}^{r_a} dr \sin\theta d\theta d\phi \frac{r^2 E^3(r,T)}{ \left(1+ \frac{2r^2 H(T,r) H_0 }{1-  H_0^2 r^2  } - \frac{H_0^2 r^2 \left(1-r^2 H^2(T,r)\right)}{\left(1- H_0^2 r^2\right)^2 }  \right)^{1/2}\, \left(1-r^2H^2 \right)^{3/2}}
\label{e32} 
\end{eqnarray}
To the best of our knowledge, the above integral cannot be evaluated analytically. However, we may easily estimate it and argue in favour of the area scaling near the Hubble horizon as follows. Since the $(1-H^2(T,r) r^2)$ term in the denominator is divergent near the Hubble horizon and the integration is defined on a constant time hypersurface,  we may safely pull the energy function out of the integral as earlier.
Also we can define the time dependent cut-off near the Hubble horizon, 
$$L_P(T)=\int_{r_C(T)-L_a}^{r_C(T)}\frac{dr}{\sqrt{1-r^2 H^2(T,r)}}$$
Now, if we are far away from the de Sitter limit, $H_0 \ll H(T,r)$ we may pull out the $\sqrt{g_{rr}}$ term off the integral as well. Putting these all in together,
and performing the trivial angular integration, we obtain
\begin{eqnarray}
P(E(T))\approx\frac{4\pi A_{\perp }(T) E^3(T)/3}{\sqrt{1+ \frac{2H_0/H}{1-H_0^2/H^2}} } \int_{r_b}^{r_a}\frac{dr}{ \left(1-r^2H^2 \right)^{3/2}}
\label{e33} 
\end{eqnarray}
Even though it is not similar to what we obtain for the case of the Killing horizon, Eq.~(\ref{e21}),  an area scaling is manifest. In fact we note that the quadratic divergence in the denominator for the Killing horizon is `tamed' here a little bit -- indicating lesser value of probability function compared to a Killing horizon of same area.  Perhaps This should be attributed to the fact that a Hubble horizon cannot hide information completely -- modes can re-enter through them. 

%%%%%%%%%%%%%%%%%
\section{Summary and discussions}
%%%%%%%%%%%%%%%%%%%%%%%%%%%%%%
\noindent
 It was shown, a few years back in~\cite{Kolekar:2010py}, that when a box of ideal gas approaches the Killing horizon of a static spherically symmetric spacetime, the entropy of the gas develops an area dependence. In this work, we have extended this result for non-static spacetimes with a positive cosmological constant, like the stationary axisymmetric and the cosmological spacetimes. 

For the stationary axisymmetric spacetimes, the imposition of Planck length cut-off in the near horizon divergent integral requires care and we have accomplished this by using the norm of the timelike, hypersurface orthogonal, vector field $\chi^a$. The final result in Eq.~(\ref{es34}) is exactly the same as that of the static and spherically symmetric spacetime.  We also have provided an expression for generic spacetime dimensions in Eq.~(\ref{es37}). These results hold in an equal footing for both black hole and the cosmological event horizons. We note that the techniques developed in~\cite{Bhattacharya:2015mja} (and references therein) were very essential in this computation, as the coordinate $R$ defined in Eq.~(\ref{es15}) was very crucial in defining the Planck length cut off near the horizon, irrespective of the path of the box.

In the case of cosmological spacetimes evolving to the (Schwarzschild-) de Sitter, we first built a suitable coordinate system to deal with the Hubble horizon, Eqs.~(\ref{e8}),~(\ref{e11}). When the spacetime is close to the de Sitter, we have explicitly evaluated the leading order expression of the entropy of the gas when it is close to the Hubble horizon and have shown that it is formally similar to that of the de Sitter. For a generic cosmological   spacetime, also, we also have formally demonstrated the area scaling when the box is close to the Hubble horizon.

It would be highly interesting to apply our methods to deal with various non-static spacetimes in other approaches as well, such as the so called brick wall formalism~\cite{'thooft}. We hope to return to this issue in the near future.

\vskip 1cm

%%%%%%%%%%%%%%%%%%%%%%%%%%%%%%%%%%%%%%%%%%%%%%%%%%%%%%%%%%%%%%%%%%%%%%%%%%%%%%%%%%%%%%%%%%%%%%%%%%%%%%%%%
\section*{Acknowledgement}
\noindent 
Majority of S.B.'s work was done when he was a post doctoral fellow at IUCAA, Pune, India. Research of S.C. is supported by the SERB-NPDF grant (PDF/2016/001589) from DST, Government of India. Research of T.~Padmanabhan is partially supported by J.~C.~Bose Research Grant, DST, Government of India. 
%%%%%%%%%%%%%%%%%%%%%%%%%%%%%%%%%%%%%%%%%%%%%%%%%%%%%%%%%%%%%%%%%%%%%%%%%%%%%%%%%%%

%%%%%%%%%%%%%%%%%%%%%%%%%%%%%%%%%%%%%%%%%%%%%%%%%%%%%%%%%%%%%%%%%%%%%%%%%%%

\end{document}